\documentclass[pra,twocolumn,preprintnumbers,amsmath,superscriptaddress]{revtex4}
\usepackage{graphicx}
\usepackage{epstopdf}
\usepackage{dcolumn}
\usepackage{bm}

\begin{document}

\title{Sagnac interferometry with coherent vortex superposition states in exciton-polariton condensates}

\author{Frederick Ira Moxley III}
\affiliation{Hearne Institute for Theoretical Physics, Department of Physics \& Astronomy, Louisiana State University, Baton Rouge, Louisiana 70803-4001, USA}

\author{Jonathan P. Dowling}
\affiliation{Hearne Institute for Theoretical Physics, Department of Physics \& Astronomy, Louisiana State University, Baton Rouge, Louisiana 70803-4001, USA}

\author{Weizhong Dai}
\affiliation{College of Engineering \& Science, Louisiana Tech University,
Ruston, LA 71272, USA}

\author{Tim Byrnes}
\affiliation{New York University, 1555 Century Ave, Pudong, Shanghai 200122, China}
\affiliation{NYU-ECNU Institute of Physics at NYU Shanghai, 3663 Zhongshan Road North, Shanghai 200062, China}
\affiliation{National Institute of Informatics, 2-1-2 Hitotsubashi, Chiyoda-ku, Tokyo 101-8430, Japan}

\date{\today}

\begin{abstract}
We investigate prospects of using counter-rotating vortex superposition
states in non-equilibrium exciton-polariton Bose-Einstein condensates for the purposes of Sagnac interferometry. We first investigate the stability of vortex-antivortex superposition states, and show that they survive at steady-state in a variety of configurations. Counter-rotating vortex superpositions are of potential interest to gyroscope and seismometer applications for detecting rotations. Methods of improving the sensitivity are investigated by targeting high momentum states via metastable condensation, and the application of periodic lattices.  The sensitivity of the polariton gyroscope is compared to its optical and atomic counterparts.  Due to the large interferometer areas in optical systems and small de Broglie wavelengths for atomic BECs, the sensitivity per detected photon is found to be considerably less for the polariton gyroscope than with competing methods.  However, polariton gyroscopes have an advantage over atomic BECs in a high signal-to-noise ratio, and have other practical advantages such as room-temperature operation and robust design.  We estimate that the final sensitivities including signal-to-noise aspects are competitive with existing methods.  
\end{abstract}

\maketitle




\section{Introduction}

Atomic Bose-Einstein condensation (BEC) was first achieved in 1995 \cite
{Anderson,Ketterle}, which inspired a massive worldwide research effort into
investigating its properties. Today, atomic BECs are being considered for
applications in a variety of fields, such as quantum simulation \cite{Bloch}%
, quantum information \cite{Byrnes12,Byrnes13}, and quantum metrology \cite
{Schumm}. Recently, it has become apparent that BECs can be made in a
variety of different systems. In particular, a BEC of quasi-particle
excitations in semiconductor systems consisting of exciton-polaritons have been
observed \cite{Deng,Byrnes14}. Exciton-polariton BECs are formed in
semiconductor microcavities \cite{Byrnes132}, or planar dielectric
Fabry-P\'{e}rot optical cavities \cite{Plumhof}, for example, and can be described as
half-matter, half-light excitations that are typically created using laser injection
methods \cite{Masumoto,Jacqmin}. Exciton-polariton BECs are a
non-equilibrium phenomenon that can be achieved at room temperatures for suitable materials that support the polaritons \cite{Houdre}. In materials where polaritons can be operated at room temperature, interactions tend to be weaker, however more recently the nonlinearities have been observed to be significant \cite{Ge13}. Despite the non-equilibrium nature of the
polariton BEC, it shares many of the properties of an atomic BEC, such as
superfluidity, stable vortex formation, and high spatial coherence. 

Although
it is now commonplace to realize atomic BECs, typically they need to be
cooled to very low temperatures in the region of $\sim 100$ nK. Even for
high-temperature superconductors, which may be viewed as a type of charged
condensation \cite{Keeling}, the highest critical temperature attained has
been $138$ K \cite{Lesne}. On the other hand, exciton-polaritons (or
polaritons for simplicity) typically have very high 
critical temperatures, due to their exceedingly light effective mass (i.e., $%
\sim 10^{-5}$ times the bare electron mass). Consequently, this opens the
door for new quantum technologies based on room-temperature superfluidity
(i.e., ``polaritonics'') \cite{Medard}, and avenues for novel light sources 
\cite{Byrnes132}.

One particular proposal based on BECs that has gathered recent interest is the matter wave 
analogue of the Sagnac interferometer, also called a ``quantum
gyroscope.'' An intuitive way to view these are as a neutral species version
of a superconducting quantum interference device (SQUID) \cite{Franchetti12}. SQUIDs are
well-known to be the most sensitive magnetometers, with capabilities of
detecting small magnetic fields in the region of $5\times 10^{-18}$ T \cite
{Clarke}. Due to the close analogy of a magnetic field of a charged particle
and rotation in the Schr\"{o}dinger equation, one may envision a device
based on BECs that is highly sensitive to rotations, instead of magnetic
fields. For this reason, ultracold atomic gases have been considered to be
an outstanding candidate physical system for realizing such a Sagnac
interferometer \cite{Riedel,Okulov}. One advantage that atomic systems possess is
the exquisite experimental control that is achievable, particularly for
creating vortices in BECs \cite{Maucher}. Moreover, these ultracold atomic
gases are highly coherent \cite{Yin}, meaning that they are particularly
suited for interferometry. However, from a practical perspective, one of the
drawbacks of technologies that utilize ultracold atomic gases is that in
order to achieve BEC, experimental conditions need to be cooled to
nanokelvin temperatures. This can prove quite expensive, and difficult to
achieve in a compact device design. In order to overcome this challenge, a
natural extension of the idea is to use exciton-polariton BECs \cite{Gu,Sturm},
which may condense at room temperatures \cite{Lerario,Dall}. This offers
a potential new configuration for achieving a novel Sagnac interferometer 
\cite{Gross}. In this paper we investigate in detail the possibility of such a ``polariton 
quantum gyroscope.''  The type of novel measurement configuration investigated herein has
the potential for providing a highly compact and stable Sagnac
interferometer \cite{Borde}, which can also be used as a seismometer, 
for example. Future applications of quantum gyroscopes
include the detection of earthquakes \cite{Michel}, gravitational waves \cite
{Dickerson,Sun}, sidereal time \cite{Gasparini}, inertial sensing \cite
{Edwards} capabilities, and even the rotational rate of our universe \cite{Delgado}.

Such an exciton-polariton gyroscope would be a hybrid optical-matter wave
Sagnac interferometer \cite{Dowling98}, as polaritons themselves consist of
both photons and excitons. In our scheme, one of the necessary ingredients for this
interferometer will be the optical excitation of vortices \cite{Nandi} in
the polariton BEC. The direct excitation of vortices by optical means is a
convenient way of producing the required vortex superposition state in the condensate, as
techniques to generate the optical counterparts are well-established. Here, it should be pointed out that the counter-rotating vortex superpositon state differs from the vortex-antivortex pair such as those observed in Ref. \cite{Roumpos}. The
investigation of various vortex states in polariton BECs has continued to
gain much interest, with several works examining disorder-induced vortices 
\cite{Grosso}, vortex-antivortex pairs \cite{Roumpos,Molina-Terriza,Dall},
and vortex ring creation via rotation of Gaussian laser beams \cite
{Yakimenko}, to name a few. One of the most efficient methods of vortex
creation is using Laguerre-Gauss optical states which carry a well-defined
external orbital angular momentum (OAM) \cite{Marzlin,Krizhanovskii,Sanvitto10}.
However, for Sagnac interferometry, a superposition of a vortex and an
antivortex is required, which has not been achieved in polariton BECs to
date to our knowledge. While the vortex-antivortex superposition has been
achieved in atomic systems \cite{Wright09}, this remains an outstanding
problem for polaritons, and is interesting from both a fundamental and
applications perspective. 

The purpose of this article is to investigate the prospects of angular momentum 
superposition states in exciton-polariton BECs for potential applications
in Sagnac interferometry.  In Sec. \ref{sec:imprint} we present and analyze a 
scheme for generating vortex-antivortex superpositions via coherent excitation with OAM states of light. 
In the scheme, the polariton BEC is initially seeded with the desired OAM superposition, 
then allowed to evolve freely via the open-dissipative Gross-Pitaevskii equation. 
This allows for the polaritons to approach their own dynamical
equilibrium state, independent of the initial state of light. We then study methods to improve the sensitivities of the polariton BEC gyroscope in Sec. \ref{sec:enhance}.  This
includes using ring geometries to target high angular momentum states, and ways of generating this using metastable condensation.  A second enhancement
scheme is discussed, where periodic lattices are imposed in order to introduce high momentum components to the wavefunction.  In Sec. \ref{sec:sagnac},
we discuss the sensitivities of the Sagnac phase for the vortex
superposition state, and compare with other existing approaches. In
particular, we will derive the Sagnac phase for the vortex superposition and ring geometry cases, and compare the performance to optical and atomic Sagnac interferometers. Finally, we will summarize our results and conclude in Sec. V.

\section{Vortex-antivortex superpositions in polariton BECs}

\label{sec:imprint}

\subsection{Initialization by Laguerre-Gauss modes}

\label{sec:exp}

\begin{figure}[tbp]
\includegraphics[width=\columnwidth]{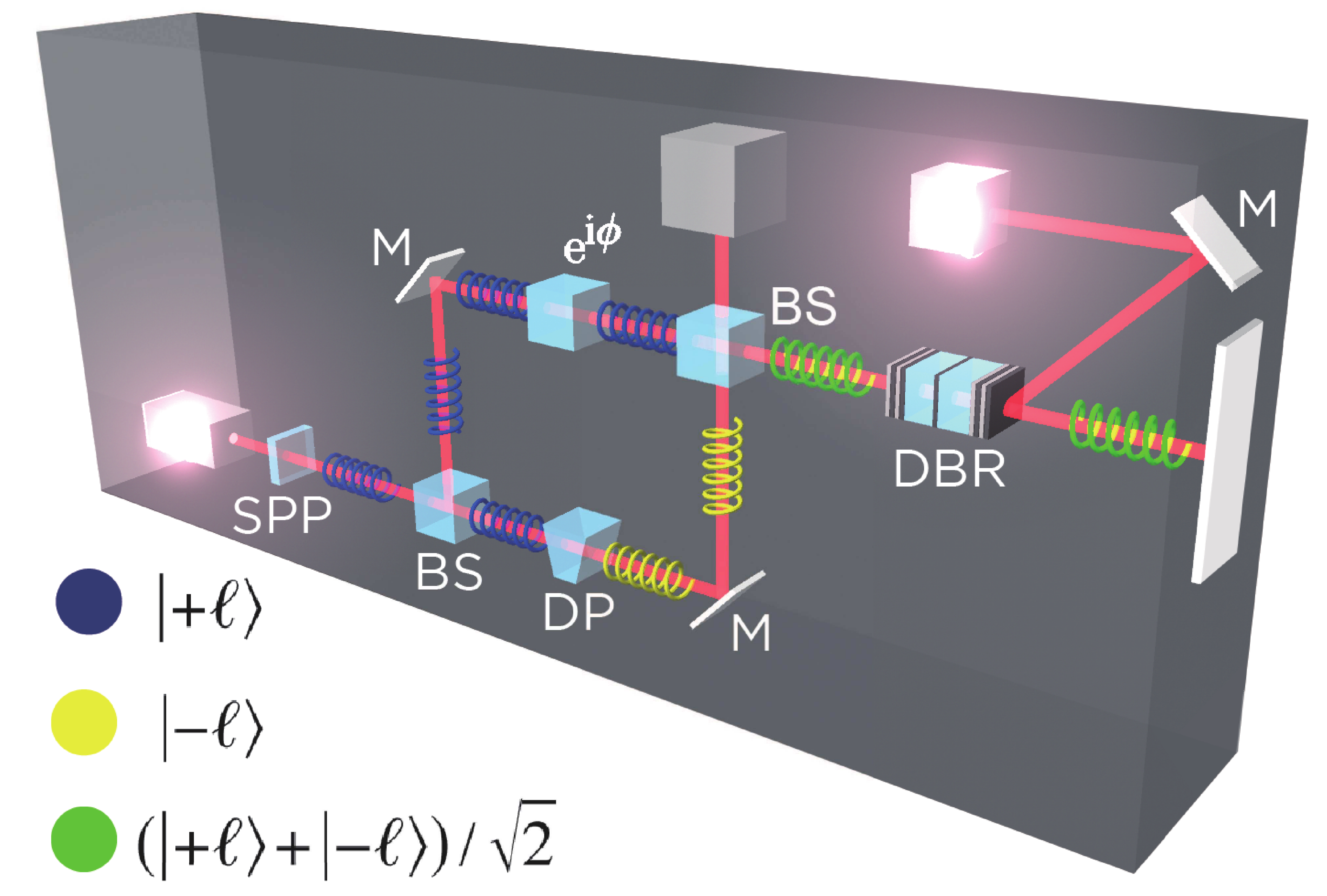}
\caption{An experimental configuration for producing vortex-antivortex
superpositions in exciton-polariton Bose-Einstein condensates. A laser
generates a TEM$_{00}$ Gaussian beam profile. This light is passed through a
spiral phase plate (SPP), generating an orbital angular momentum (OAM) state
of light. Once the OAM state of light has traveled through the Mach-Zehnder
interferometer with a Dove prism (DP) and mirrors (M), an OAM superposition state of light is
made. The resulting OAM superposition state of light can be used to generate
an interference pattern in a polariton BEC matter wave within the
distributed Bragg reflector (DBR) microcavity. This interference pattern is
imaged by a detector such as a charge-coupled device (CCD), and can then be
used to determine the Sagnac phase and therefore the rotational rate of the polariton BEC. 
}
\label{fig1}
\end{figure}

We first describe how the vortex-antivortex superposition is created in the
exciton-polariton BEC. The experimental setup for the creation of a
vortex-antivortex polariton BEC is shown in Fig. \ref{fig1}, which is similar to
the traditional optical (i.e., Mach-Zehnder) Sagnac interferometer. Unlike the traditional Mach-Zehnder scheme, our polariton BEC interferometer
is based on a principle that employs OAM superposition states of light. Our starting 
point is a source of OAM states of light, which can readily be produced by established methods, for example, by using  
spiral phase plates (SPPs).  This may be used to create OAM superpositions, which coherently excite the 
polaritons in a DBR structure (e.g., semiconductor). This produces the initial seed polariton condensate, from
which the polariton BEC is reinforced by the standard process of
exciton-polariton cooling and scattering.

Let us denote the bosonic annihilation operator for an OAM mode with angular momentum number
$ l $ and radial node number $ p $ as $a_{l,p}$  (see Appendix \ref{sec:app} for more details). Then the superposition state for the mode that we require is $\alpha a_{l,p}+\beta a_{-l,p}$, where $|\alpha |^{2}+|\beta
|^{2}=1.$ This OAM superposition state of light has two counter-rotating
components which will induce an interference pattern in the polariton BEC.
Once the characteristic interference of the OAM superposition (consisting of 
$2l$ lobes) has been transfered to the polariton BEC, the Sagnac phase can
then be determined from the angular velocity of the BEC cloud. We can obtain
an ordinary vortex state here simply by setting either $\alpha $ or $\beta =0
$ \cite{Dreismann}.

Many lasers emit light beams which approximate a Gaussian profile with  $l=p=0$. Using the TEM$_{00}$ fundamental
transverse mode of the optical resonator and channeling it through a SPP,
for example, can be used to generate OAM states of light (see Appendix \ref{sec:app}). 
Although these states are generally made with holographs, SPPs
offer a simple and convenient way to obtain a pure OAM state $a_{l,p}$ with arbitrary integer $l$. Moreover, once the pure state $a_{l,p}$ has been obtained, we can generate an OAM superposition
state. One way to accomplish this is by using a Mach-Zehnder type of configuration as
seen in Fig. \ref{fig1}, where the Dove prism is used to transform the $a_{l,p}$ state into a $a_{-l,p}$ state to obtain $%
(a_{l,p}+a_{-l,p})/\sqrt{2}$ at the real output of the Mach-Zehnder
interferometer. If desired, one may choose the first beam splitter as an $%
\left| \alpha \right| ^{2}$:$\left| \beta \right| ^{2}$ beam splitter, and
the second one as a 50:50 beam splitter to obtain an arbitrary superposition
of the type $\alpha a_{l,p}+\beta a_{-l,p}$.

Let us now consider the specific case as shown in Fig. \ref{fig1}, where
the two inputs of the first beam splitter are prepared in a coherent state
in the LG mode $a_{l,p}$ and the vacuum state respectively. Hence, 
\begin{equation}
a_{\text{IN}}=a_{l,p}.  \label{eq12}
\end{equation}
Assuming that the first beam splitter being used is a symmetric 50:50 beam
splitter, 
the OAM state of light has transformed like 
\begin{equation}
a_{\text{IN}}=a_{l,p}\mapsto \frac{1}{\sqrt{2}}\left( b_{l,p}+c_{l,p}\right)
,  \label{eq14}
\end{equation}
where $b_{l,p},c_{l,p}$ denotes the LG modes in the two arms of the
interferometer. The phase shift in the upper arm of the Mach-Zehnder
interferometer will cause the coherent amplitude to replaced by $%
a_{l,p}\mapsto e^{i\phi }a_{l,p}$, and the Dove prism in the lower arm of
the Mach-Zehnder interferometer will cause the coherent amplitude to be
replaced by $b_{l,p}\mapsto b_{-l,p}$. Hence, the OAM state of light seen
in Eq. (\ref{eq14}) becomes 
\begin{equation}
\frac{1}{\sqrt{2}}\left( b_{l,p}+c_{l,p}\right) \mapsto \frac{1}{\sqrt{2}}%
\left( e^{i\phi }b_{l,p}+c_{-l,p}\right) .  \label{eq15}
\end{equation}
Finally, neglecting the imaginary output of the interferometer, the second
beam splitter reverses the transformation and produces the OAM superposition
state of light, written as 
\begin{equation}
a_{\text{OUT}}=\frac{1}{\sqrt{2}}\left( a_{l,p}+e^{i\phi }a_{-l,p}\right) ,
\label{eq16}
\end{equation}
which is our desired output. The superposition of $a_{l,p}$ and $a_{-l,p}$
states in Eq. (\ref{eq16}) can be detected using many different techniques,
which typically includes imaging the interference pattern of the light beams
which constitute the OAM superposition state. In particular once the
polariton BEC is induced, one may use standard spatial and momentum space imaging methods to
analyze the coherent condensate state in the polariton microcavity \cite
{Byrnes14}. 

\subsection{Exciton-polariton BEC dynamics}

The laser injection methods of the previous section can be used to
resonantly excite the polariton BEC. The phase of the light is directly
imprinted on the polariton BEC system, hence the injected polaritons simply
follow the same phase relation as the injected light. In this configuration
alone, there is no difference of using the OAM superposition states of light
to the polariton BEC, as the polaritons inherit their vortex nature from the
light. However, once the initial coherent excitation is induced, the OAM
superposition is turned off, and the microcavity system is excited by
conventional polariton pumping. This can be performed by an off-resonant
pumping where the excitons have a much higher energy than the polaritons, 
or by pumping
at a high in-plane momenta (see Figure 1b in Ref. \cite{Byrnes14}). Either
method creates a large population of uncondensed polaritons from which the
polariton BEC is replenished. It is important to note that apart form the
initial seed polariton BEC which is induced coherently, the remaining
population thereafter is not pumped with the OAM superposition state. Hence
in the second phase where the conventional polariton BEC pumping is used,
any OAM superposition state is a genuine steady state of the polariton BEC,
and possesses its own dynamics.

While both vortices and vortex-antivortex pairs have been shown theoretically and
experimentally to be stable configurations of polariton
BECs \cite{Tosi,Roumpos}, whether or not a counter-rotating vortex superposition is a stable configuration is a non-trivial
problem. Such vortex-antivortex superpositions have a more complex phase
relation than standard vortex configurations, and have the possibility to
spontaneously separate spatially, or relax to a zero momentum state. To
investigate this, we have studied the vortex-antivortex superposition in a
variety of different circumstances to evaluate its stability. 

The
spatiotemporal evolution of the non-equilibrium exciton-polariton system is
described by the open-dissipative Gross-Pitaevskii (dGP) equation,
expressed as \cite{Keeling,Lagoudakis} 
\begin{align}
i\frac{\partial \psi (\mathbf{x},t)}{\partial t}& =\Big[ -\frac{\hbar }{2m}%
\nabla ^{2}+V_{ext}(\mathbf{x})+g\left| \psi (\mathbf{x},t)\right| ^{2}  \nonumber \\
& +\frac{i}{2}(P(\mathbf{x})-\gamma -\eta \left| \psi (\mathbf{x},t)\right| ^{2})\Big] %
\psi (\mathbf{x},t),  \label{eq17}
\end{align}
where $\psi (\mathbf{x},t)$ is a complex valued condensate order parameter, $%
g$ is the polariton-polariton interaction constant, $m$ is the polariton
mass, $V_{ext}(\mathbf{x})$ is a spatially dependent trapping potential energy (see Figure 5 in Ref. \cite{Byrnes14}), $P(\mathbf{x})$ is a spatially dependent
pumping rate, $\gamma $ is the polariton loss rate, and $\eta $ is the gain
saturation \cite{Wouters07,Wouters10}. Throughout this paper we will assume that the time scale is in units of 
$ 2 m a/ \hbar $, where $ a $ is the length scale set by the experiment (e.g., size of trapping potential or pump spot diameter). 

We use the Generalized Finite-Difference Time-Domain (G-FDTD) scheme to solve the dGP equation. 
G-FDTD is an explicit method that permits an accurate solution with simple computation for solving Eq. (%
\ref{eq17}) in multiple dimensions. To this end, the function $\psi (\mathbf{x}%
,t)$ is first split into real and imaginary components resulting in two
coupled equations. The real and imaginary components are then approximated
using higher-order Taylor series expansions in time, where the derivatives
in time are then substituted into the derivatives in space via the coupled
equations. Finally, the derivatives in space are approximated using
higher-order finite difference methods. The G-FDTD has been successfully
applied for solving both linear and nonlinear Schr\"{o}dinger equations \cite
{Moxley13,Moxley15}.

\subsection{Stability of vortex-antivortex superpositions}
\label{sec:stability}

\begin{figure}[tbp]
\includegraphics[width=\columnwidth]{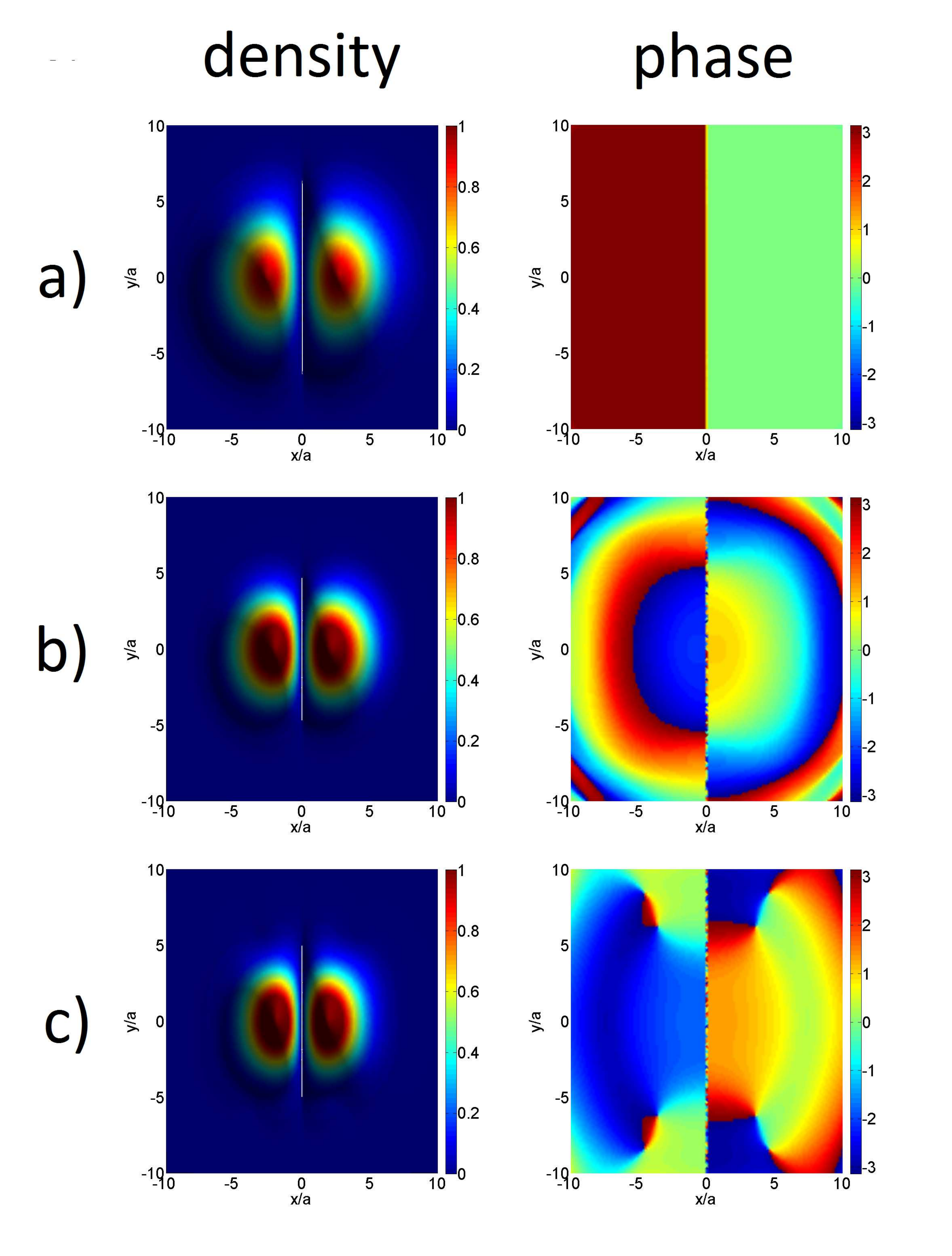}
\caption{Time evolution of a polariton BEC initially seeded with a vortex-antivortex superposition for a  flat potential $ V_{ext}(\mathbf{x}) = 0 $. 
Timeframes for the  (a) initial state $t = 0$; (b) intermediate state $ t = 5 $; and (c) steady-state $ t = 10 $ are shown. Parameters used were 
$P(x,y) = P_0 e^{-(r/r_0)^2}$, where $P_0= 2$, $r_0 =5.35$ , and $g= \gamma = \eta = 1$. Timescales are in units of $ 2m a^2/\hbar $ where $ a $ is the lengthscale unit. }
\label{fig3old}
\end{figure}

We now examine the stability of the vortex-antivortex superposition under the dynamical evolution of the dGP equation. 
In Fig. \ref{fig3old}, we show the spatiotemporal evolution of the non-equilibrium polariton 
BEC vortex superposition state in a flat potential landscape $ V_{ext} =0 $. The initial condition was chosen as 
\begin{equation}
\psi(x,y,t=0) = \frac{P(x,y)-\gamma }{\sqrt{2} \eta } (u_{l=1,p=0}^{\text{LG}}+u_{l=-1,p=0}^{\text{LG}%
}).  \label{eq21}
\end{equation}
where $ u_{l,p}^{\text{LG}} $ are the spatial Laguerre-Gauss wavefunctions defined in Appendix \ref{sec:app}. The factor of $  \frac{P(x,y)-\gamma }{\eta }$ approximates the equilibrium density which the dGP equation converges to. For densities near equilibrium 
\begin{equation}
|\psi (x,y,t)|^{2}=\frac{P(x,y)-\gamma }{\eta },  \label{eq19}
\end{equation}
the imaginary terms on the right hand side of Eq. (\ref{eq17}) vanish and we
may obtain a zeroth order estimate for solutions \cite{Pitaevskii}. Starting from this approximation,
the numerics should converge towards the true solution for the dGP equation, which
will in general be different from the standard (conservative) Gross-Pitaevskii solution. Throughout this
study, we assumed that the wavefunction in the $z$-direction has an
insignificant effect on the vortex dynamics in the $x$-$y$ plane.
In the limit $P=\gamma =\eta =0$, the condensate is conservative (i.e., the particle number is
fixed and there is nothing entering or leaving the condensate). With $%
P-\gamma ,\eta >0$, the condensate has both leakage and gain, but stability
is maintained such that there is no net loss.

From the initial condition given by Eq. (\ref{eq21}), the system is time-evolved under Eq. (\ref{eq17}) until a steady state is reached. Fig. \ref{fig3old} illustrates
the condensate density distribution $|\psi (x,y,t)|^{2}$ of the
two-dimensional vortex non-equilibrium condensate in a flat
potential at various times, and also illustrates the phase (i.e., the argument of $ \psi (x,y,t) $ ) of the two-dimensional vortex non-equilibrium condensate in a
flat potential at various times.  Upon examination of Fig. \ref{fig3old}, 
we can see that our counter-rotating vortex-antivortex superposition state is 
coherent and stable as it is generated in the distributed Bragg reflector (DBR) microcavity 
outlined in Fig. \ref{fig1}.  Here, it should be pointed out that the vortex-antivortex superposition has the characteristic phase dependence of the lobes being out of phase with each other by a factor of $ \pi $.  The system quickly reaches steady state and little change is seen beyond 
$ t = 10 $. The stability of the counter-rotating vortex-antivortex superposition can be attributed to the same reason as for 
atomic BECs: both the vortex and antivortex require a density defect due to the phase velocity diverging at the vortex
core.  Thus it is energetically favorable for both to share the same vortex core, effectively pinning them in the 
same location.  

\begin{figure}[tbp]
\includegraphics[width=\columnwidth]{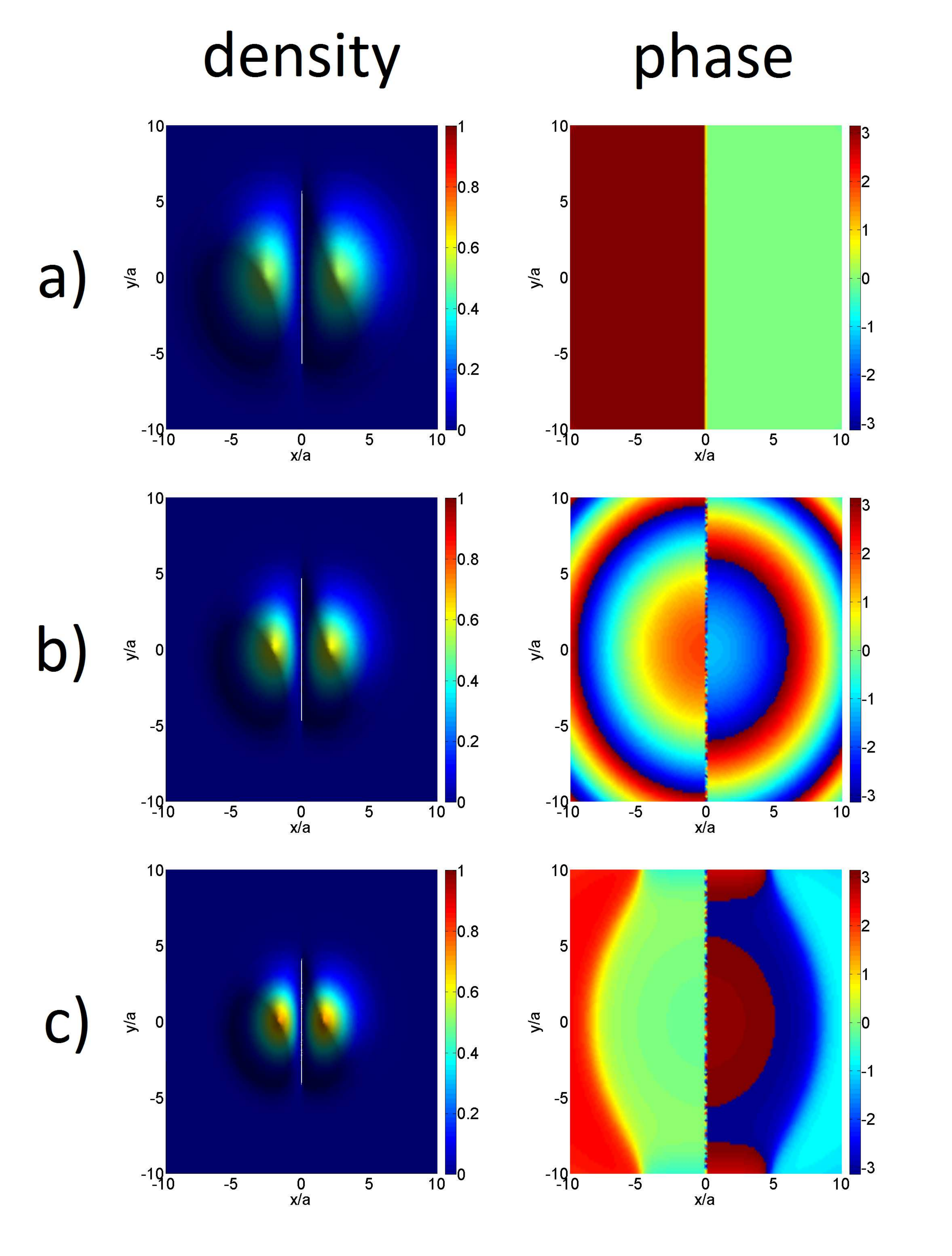}
\caption{Time evolution of a polariton BEC initially seeded with a vortex-antivortex superposition for a  flat potential $ V_{ext}(\mathbf{x}) = 0 $. 
Timeframes for the  (a) initial state $t = 0$; (b) intermediate state $ t = 5 $; and (c) steady-state $ t = 10 $ are shown. Timescales are in units of $ \hbar/meV $. The following parameters were used in the simulations: $ \hbar g = 0.05$,  $ \hbar \gamma = 1.0$, and $ \hbar \eta = 0.1$, where energies are in $ meV $.  }
\label{fig3}
\end{figure}

In Fig. \ref{fig3} we also simulate the vortex-antivortex superposition for experimental parameters as given in Refs. \cite{Manni11,Wouters102,Ge13}.  We see that generally the results are unchanged and the interference lobes are stable under different parameters.  As the vortex superposition state consists of lobes, the Sagnac 
phase can be determined from the rotation of the lobes about an axis of rotation. This will be discussed in more detail in Sec. \ref{sec:sagnac}.   

\begin{figure}[tbp]
\includegraphics[width=\columnwidth]{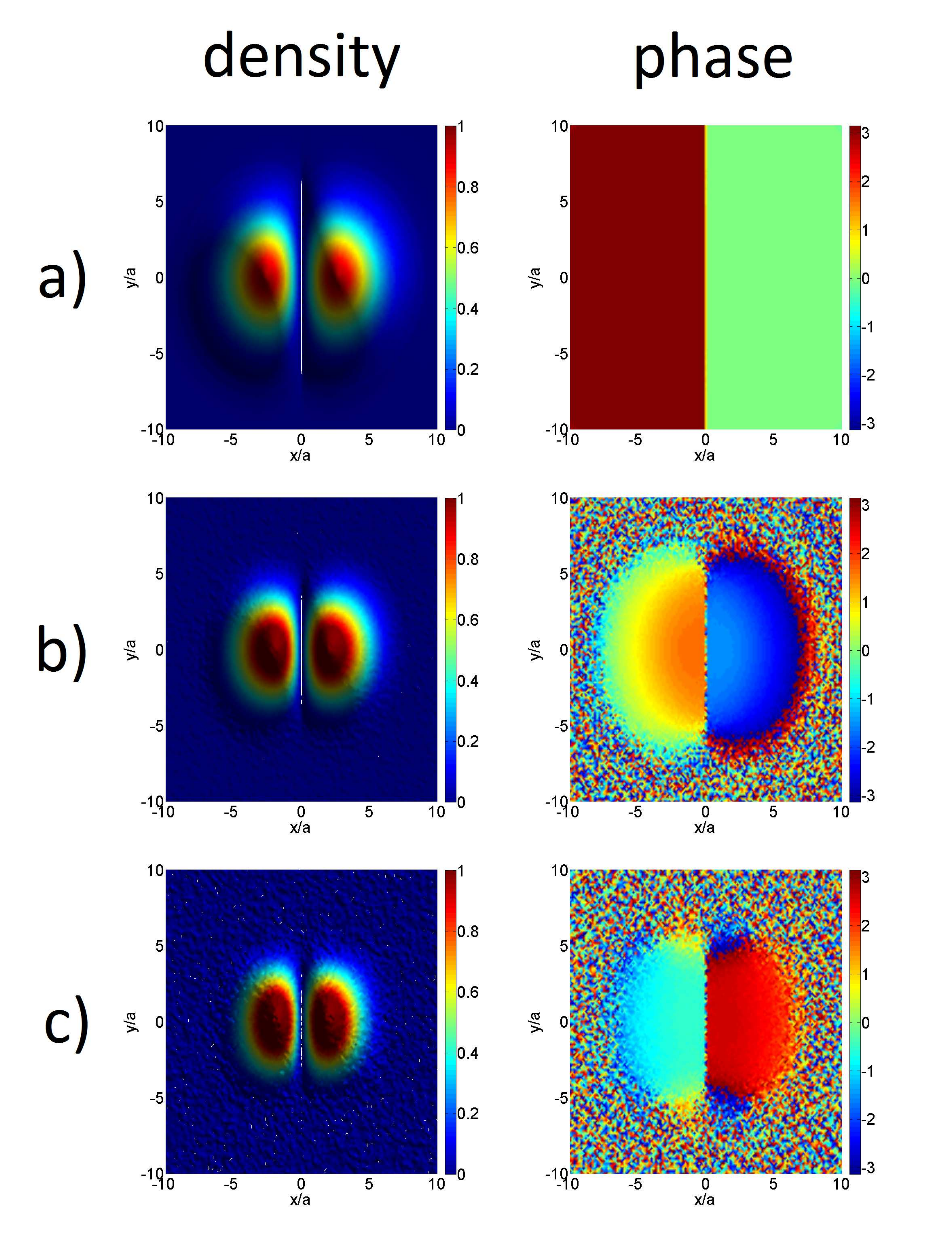}
\caption{Time evolution of a polariton BEC initially seeded with a vortex-antivortex superposition for a disordered potential. 
Timeframes for the  (a) initial state $t = 0$; (b) intermediate state $ t = 5 $; and (c) steady-state $ t = 10 $ are shown. Parameters used were 
$P(x,y) = P_0 e^{-(r/r_0)^2}$, where $P_0= 2$, $r_0 =5.35$ , and $g= \gamma = \eta = 1$. Timescales are in units of $ 2m a^2/\hbar $ where $ a $ is the lengthscale unit.}
\label{fig5old}
\end{figure}

\begin{figure}[tbp]
\includegraphics[width=\columnwidth]{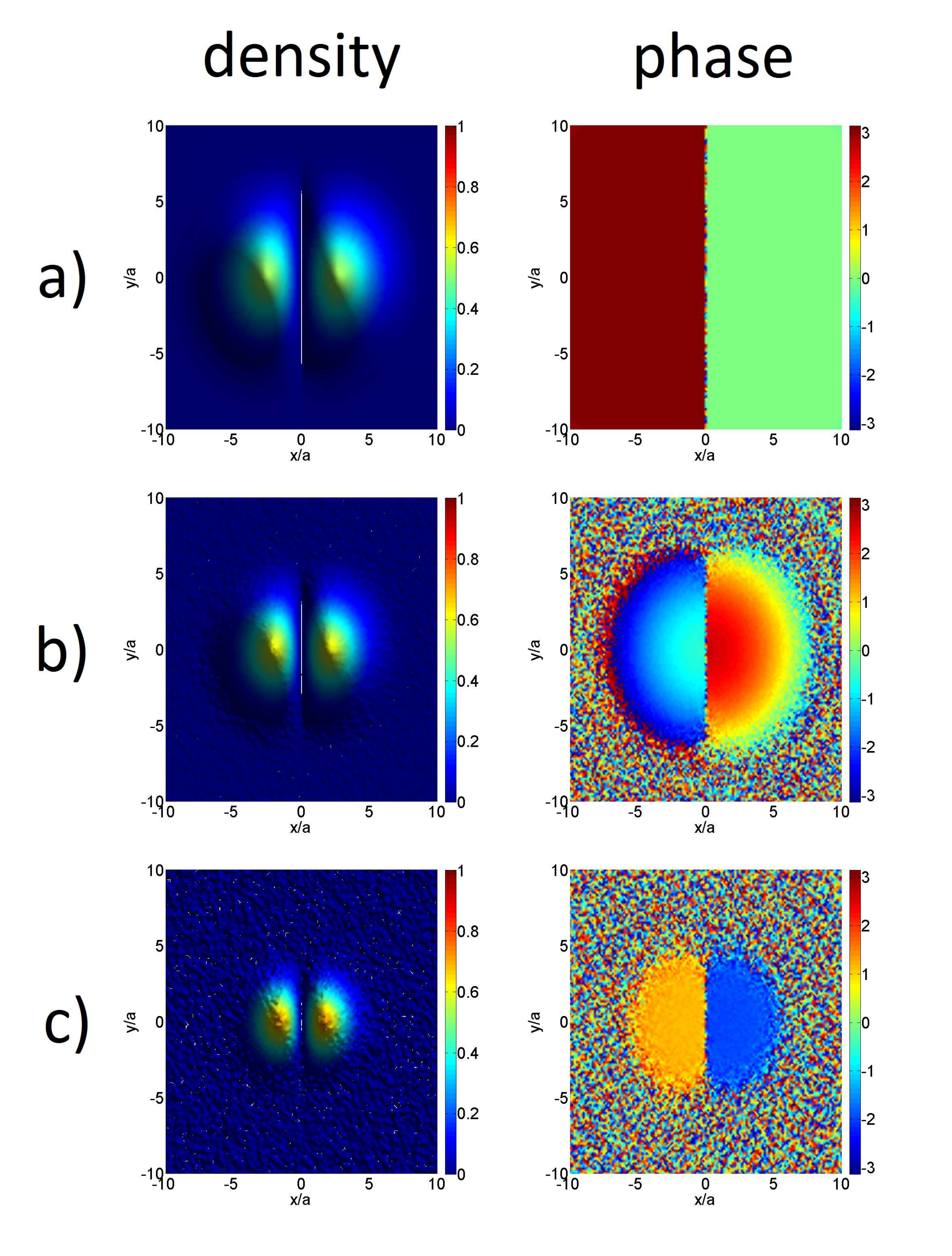}
\caption{Time evolution of a polariton BEC initially seeded with a vortex-antivortex superposition for a disordered potential. 
Timeframes for the  (a) initial state $t = 0$; (b) intermediate state $ t = 5 $; and (c) steady-state $ t = 10 $ are shown. Timescales are in units of $ \hbar/meV $. The following parameters were used in the simulations: $ \hbar g = 0.05$,  $ \hbar \gamma = 1.0$, and $ \hbar \eta = 0.1$, where energies are in $ meV $. }
\label{fig5}
\end{figure}

While Fig. \ref{fig3old} suggests that the vortex-antivortex superposition is stable, a realistic 
semiconductor system does not have a perfectly smooth potential $ V_{ext} =0 $.  There are typically
unavoidable sources of disorder, which may potentially affect the stability of the superposition 
states.  To investigate this, we study the evolution of the vortex-antivortex superposition state in a disordered potential landscape.  
Upon examination of Fig. \ref{fig5old}, we can see that in the presence of a disordered potential, our vortex superposition state is 
coherent and stable as it is generated in the distributed Bragg reflector (DBR) microcavity 
outlined in Fig. \ref{fig1}. Fig. \ref{fig5old} also illustrates the phase of the two-dimensional vortex non-equilibrium condensate in a 
disordered potential, which shows in general the same behavior as the flat potential case. We have tried this for several disorder types and strengths, 
and find that the vortex-antivortex superposition is stable in all cases.  Again, upon examination of Fig. \ref{fig5}, we can see that with experimental parameters taken from Refs. \cite{Manni11,Wouters102,Ge13} the results are largely unchanged.

\section{Sensitivity enhancements for polariton gyroscopy}
\label{sec:enhance}

The results of Sec. \ref{sec:imprint} suggest that counter-rotating vortex-antivortex superpositions can be coherently excited and are 
stable in polariton BECs.  In the context of gyroscopy, rotations are detected by changes in the lobe interference pattern as found in Figs. \ref{fig3old} - \ref{fig:ring}. Thus for sensitive detections one would like the lobes to be as sharply defined as possible, to be able to discriminate between the original and rotated interference patterns.  For vortex-antivortex superpositions in polaritons, this can be achieved using large angular momentum states $ l $ (see Fig. 2 of Ref. \cite{Fickler12}). While this is possible in principle, there are practical limits on the high angular momentum Laguerre-Gauss modes that can be excited.  Hence alternative strategies for sensitivity boosts are desirable for the polariton gyroscope.

\subsection{Metastable condensation in ring geometries}
\label{sec:meta}

In general, higher sensitivities are achieved by the use of high momentum states due to their short wavelength, allowing for high resolutions. A variant of the vortex-antivortex superposition state is to use ring geometries, where the polariton condensate is placed in a circular narrow channel potential. In a BEC, high angular momentum states are energetically disfavored, and typically multi-quantized vortices break up into many $ l = \pm 1 $ vortices \cite{Pitaevskii}.  This can be understood to result from the high momentum states that are near the vortex core which have a rapid phase variation.  By eliminating the vortex core with the use of a ring potential, high angular momentum states become ultra-stable, which is beneficial from a sensitivity point of view.

\begin{figure}[tbp]
\includegraphics[width=\columnwidth]{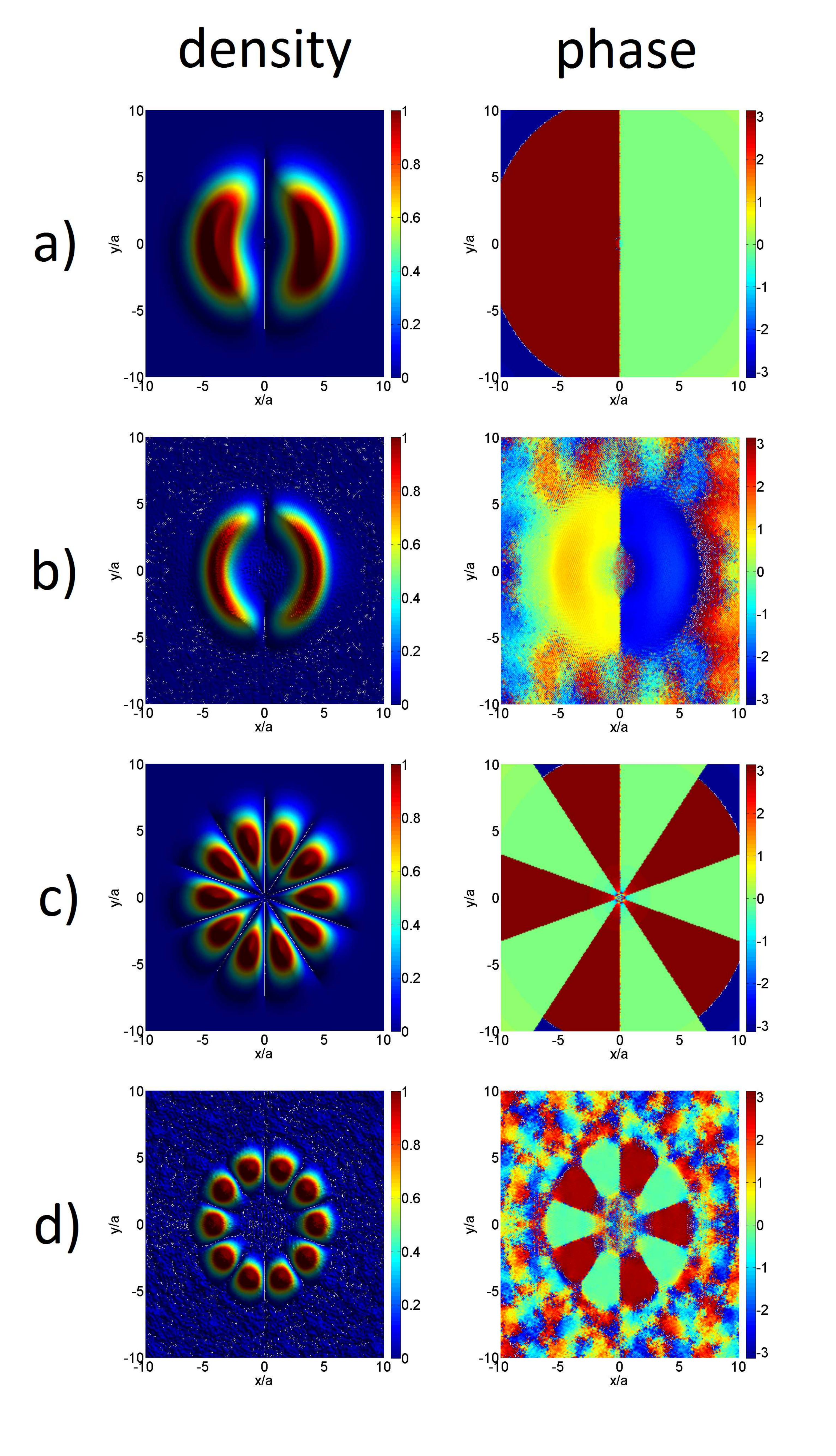}
\caption{Counterpropagating currents in ring geometries. Cases shown are (a) $ l = \pm 1 $, $ t = 0 $; (b) $ l = \pm 1 $, $ t = 10 $; (c) $ l = \pm 5 $, $ t = 0 $; (b) $ l = \pm 5 $, $ t = 10 $. Parameters used are $ \gamma = \eta = V_0 = g = 1 $, $ r_{min} = 5 $. A spatially dependent pump $ P(x,y) = P_0 | \psi_l (r,\phi) |^2 $, with $ P_0 = 2 $ is used.  Timescales are in units of $ 2m a^2/\hbar $ where $ a $ is the lengthscale unit.  }
\label{fig:ring}
\end{figure}

In Fig. \ref{fig:ring} we show the stability of counter-propagating polariton superfluid currents in ring geometries. The potential that is used is a ``mexican hat'' form
\begin{align}
V_{ext}(r) = V_0 \left( \frac{r^4}{r_{min}^4} - 2 \frac{r^2}{r_{min}^2} \right),
\end{align}
where $ r = \sqrt{x^2+y^2} $.  Expanding around the minima of the channel, one obtains an approximate wavefunction 
\begin{align}
\psi_l (r,\phi) = \exp \left[ - \frac{\sqrt{V_0}}{ r_{min}} ( r - r_{min})^2 \right] e^{-il \phi }
\end{align}
where we have added a phase variation around the ring with angular momentum $ l $.  We then start in the initial state
\begin{align}
\psi (r,\phi) = \frac{\psi_l (r,\phi) + \psi_{-l} (r,\phi)}{\sqrt{2}}
\end{align}
and evolve the system using Eq. (\ref{eq17}).  
Figure \ref{fig:ring} shows our results for evolving $ l = \pm 1 $ and $ l = \pm 5 $ superpositions.  We see that 
steady state configurations are reached with the superpositions being maintained for all times, in the same way as for the 
vortex-antivortex superpositions.   

While the above shows that with a suitable initialization of the the polariton BEC with counterpropagating currents is a stable configuration, a question is how in practice such a state would be induced.  The method of coherent excitations as discussed in Sec. \ref{sec:exp} is viable for low angular momenta superpositions, but for higher angular momenta superpositions this can be problematic \cite{Fickler12}.  An alternative method of forming the superpositions states is via metastable condensation. In Ref. \cite{Lai07} it was experimentally found that in a one-dimensional periodic lattice, polariton condensates at high momentum states could be generated.  This was attributed to the tradeoff between relaxation from high energy states and the finite decay time of the polaritons.  Here we show that it is possible to target particular momentum state superpositions using the metastable condensation dynamics and spatially dependent pumping profile.  

For large rings we may model the ring geometry as discussed above by a toroidal geometry where the $ x $ direction is the direction along the channel, and $ y $ is in the radial direction.  Then the ring geometry imposes that the wavefunction is periodic in $ x $
\begin{align}
\psi(x,y) = \psi(x+L,y)
\end{align}
where $ L $ is the circumference of the ring.  We also take for convenience that $ \psi(x,y) = \psi(x,y+L) $ although this is not important for the purposes of our illustration, and other boundary conditions could be used just as efficiently. 

Our target configuration in this case is that we have two counterpropagating momentum modes $ \pm k_0 $.  In a similar way as aforementioned, in the ideal case where only these two modes are initialized, this is a stable configuration of the polariton BEC.  To examine what happens with an imperfect initialization procedure, let us consider that we have in addition a component with zero momentum
\begin{align}
\psi(x,y,t=0) = \xi_{k=0} + \xi_{k_0} \left( e^{ik_0x} + e^{-ik_0 x} \right)  .
\end{align}
We compare two cases where the pump is spatially uniform 
\begin{align}
P(x,y) = P_0
\label{uniformpump}
\end{align}
and a spatially varying pump which has the same pumping profile as the target wavefunction density, which in this case is $ \cos^2 k_0 x $.  
\begin{align}
P(x,y) = P_0 \eta \cos^2 k_0 x + \gamma . 
\label{periodicpump} 
\end{align}
In the case of Ref. \cite{Lai07}, the spatially periodic pump modulation occurs because of the metallic deposits on the surface of the DBR which physically blocks the laser pump.  These metal deposits also have the effect of producing a periodic potential of the form
\begin{align}
V_{ext}(x,y) = V_0 \cos (2 k_0 x),  
\label{periodicpot}
\end{align}
induced by the change in the photonic boundary conditions at the surface \cite{Kim08}. The factor of 2 arises because for $ \pi $ state condensation as observed in Ref. \cite{Lai07}, the phase in the wavefunction changes by $ \pi $ between adjacent lattices.  

\begin{figure}[tbp]
\includegraphics[width=\columnwidth]{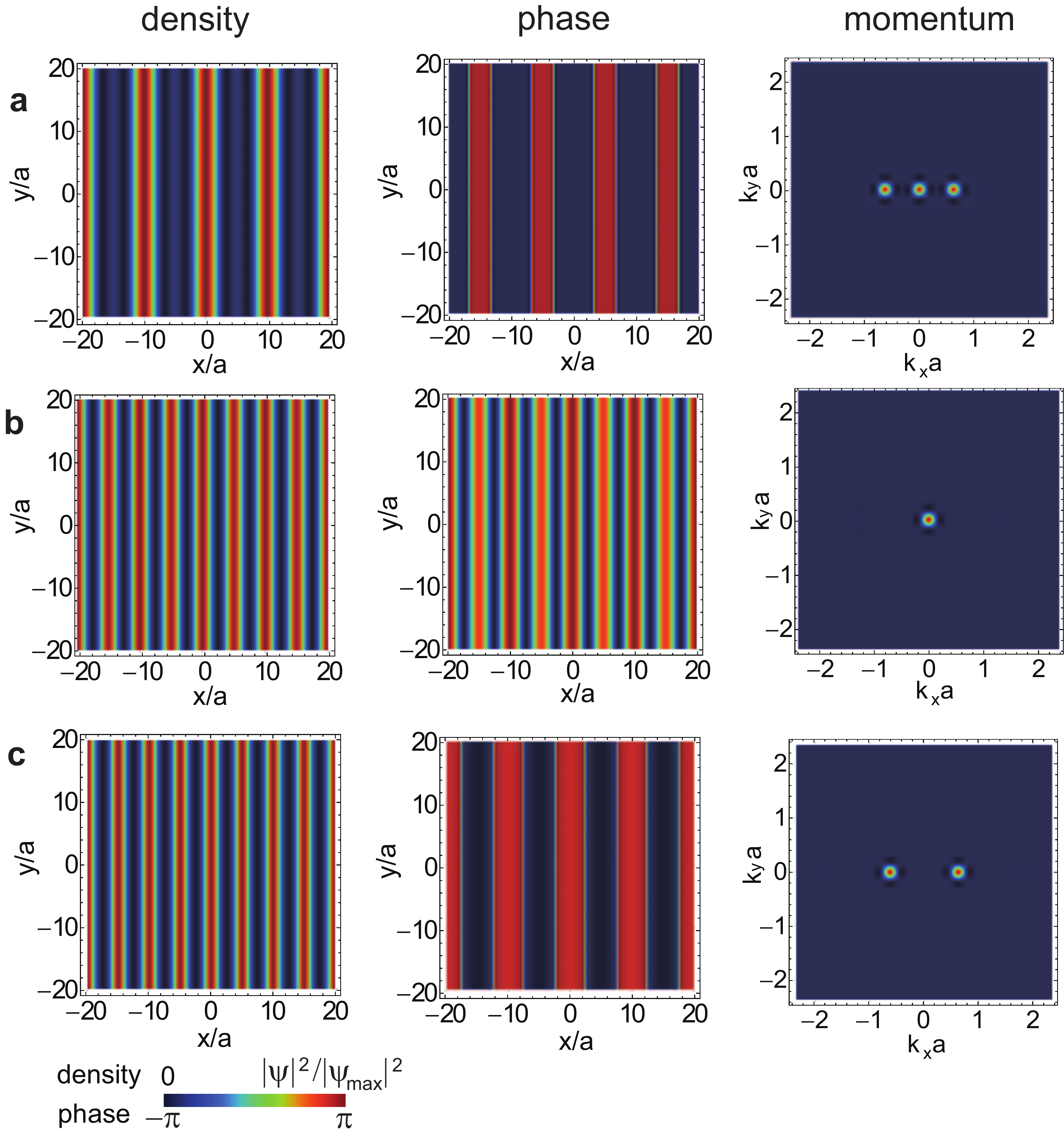}
\caption{Metastable condensation dynamics. (a) The initial state with equal populations of the three momenta $ k = 0, k_0, -k_0 $.  Time evolved steady state distributions for (b) uniform pumping Eq. (\ref{uniformpump}) and (c) periodic pumping Eq. (\ref{periodicpump}).  Parameters used are $ \gamma = \eta = V_0 = g = 1 $, $ P_0 = 2 $, $ k_0 = 2 \pi/10 $, $ \xi_{k=0} = \xi_{k=k_0} = \xi_{k=- k_0} = 1/\sqrt{3} $.   Timescales are in units of $ 2m a^2/\hbar $ where $ a $ is the lengthscale unit.  }
\label{fig:metastable}
\end{figure}

Fig. \ref{fig:metastable} shows our results. We start in a state with an equal distribution of momenta, $ \xi_{k=0} = \xi_{k=k_0} = \xi_{k=- k_0} = 1/\sqrt{3} $.  Under the uniform pumping scheme Eq. (\ref{uniformpump}), the steady state evolves towards dominantly the zero momentum mode. There is a spatial dependence to the real space wavefunction according to the potential Eq. (\ref{periodicpot}), but the phase is the same across all the sites, indicating that it is dominantly a zero momentum mode. This is explicitly seen by the Fourier space decomposition in Fig. \ref{fig:metastable}(b).  
For the periodic pumping Eq. (\ref{periodicpump}), the system evolves instead towards the momenta $ \pm k_0 $. The spatial distribution appears to be identical to the uniform pumping case, but the phase 
changes by $ \pi $ for adjacent maxima. This may be compared to the density and phase profiles in the ring geometry in Fig. \ref{fig:ring}, which shows the same variation. 

This above shows that, a spatially dependent pump may be used to target particular momentum modes with the same periodicity.  We note that without an initial seed momentum state, the pumping technique cannot generate the desired momentum modes.  The reason is that unlike the coherent excitation technique of Sec. \ref{sec:exp}, the pumping profile $ P(x,y) $ does not contain any phase information, hence cannot impose any particular phase on the condensate.  Thus some population in the desired momentum mode must be there initially, but from there the pumping can reinforce this population, and diminish the other contributions.  In practice, below the condensation threshold many momentum modes are occupied and hence as only crosses the critical pump density the system should naturally gravitate towards the target momentum states.

\subsection{Periodic potentials}

Another strategy that can be used to enhance the sensitivity is via the use of
periodic potentials on the underlying DBR lattice. The origin of this effect is the increased effective mass of the polaritons in the periodic lattice.  In general, when a periodic potential is applied to a system, the energy-momentum dispersion breaks up into bands.  Within each band, the stronger the potential that is applied, the flatter the dispersion becomes. This originates from the fact that the kinetic energy is suppressed due to the reduced tunneling between adjacent sites.  Flatter bands result in a larger effective mass of the particles.  As the de Broglie wavelength of the particles is $ \lambda = h/ m v $, where $ v $ is the velocity of the particle, a large mass gives a shorter wavelength, and thus increased sensitivities. 

\begin{figure}[tbp]
\includegraphics[width=\columnwidth]{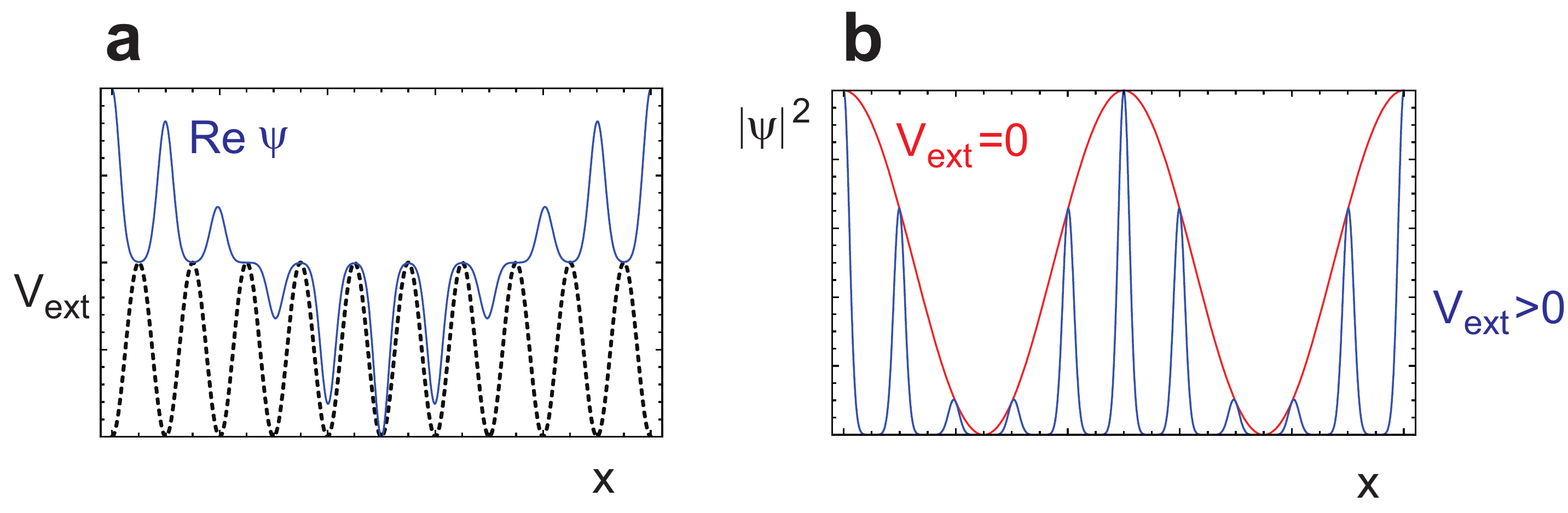}
\caption{Schematic for sensitivity enhancement effect for periodic potentials. (a) The real part of the wavefunction in a Bloch band (solid line) and the periodic potential (dashed line).  (b) The resulting interference due to two waves of opposing momentum.  Due to the spatial modulation of the Bloch wavefunction, the resolution of the interference parttern is enhanced.  }
\label{fig:periodic}
\end{figure}

A more direct physical argument which shows the increased sensitivity is shown in Fig. \ref{fig:periodic}.  Due to Bloch's theorem, any
eigenstate in a periodic potential may be written
\begin{align}
\psi_k (x) = e^{ikx} u_k (x)
\end{align}
where $ u_k (x) $ is a periodic function with the same periodicity as the crystal and $ k $ is the wavenumber.  
The envelope function $ u_k (x) $ can be decomposed into Wannier functions, which for 
strong potentials can be approximated by the localized potential around one of the potential minima in the periodic function. Thus in a periodic potential with counter-propagating modes the interference pattern is
\begin{align}
|\frac{e^{ik_0 x}+e^{-ik_0 x}}{\sqrt{2}}  u_{k_0} (x)|^2 =  \cos^2 k_0 x u^2_{k_0} (x) .
\end{align}
As illustrated in Fig. \ref{fig:periodic}, the periodic potential thus gives a finer periodicity than the $ \pm k_0 $ modes. By detecting shifts in the finer fringes this can give rise to higher sensitivities for the gyroscope. The mechanism for the increased sensitivity originates from the introduction of high momentum components in the periodic potential $ u_k (x) $. This is consistent with the more straightforward strategy of increasing $ k_0 $, except that the high momentum modes are naturally present in a periodic lattice.  For the standard approach one must inject the system with a momentum $ \pm k_0 $.

To check the stability of vortex-antivortex superpositions in periodic lattice potentials, we again perform a time evolution under Eq. (\ref{eq17}).  Specifically, we choose a Kagome or a basketweave lattice \cite
{Mekata03}. A Kagome lattice consists of interlaced triangles and exhibits a
higher degree of frustration when compared with other 2D lattices (e.g.,
square and triangular lattices) \cite{Levi07}. A Kagome lattice is
particularly interesting as in the tight-binding limit with only
nearest-neighbor tunneling the 2D Kagome lattice geometry possesses a
completely flat band. In such flat bands, the polariton condensate order
parameters become tightly localized at the potential dips \cite{Bergman08}. 
Experimentally, several schemes have been proposed to create flat bands in
the 2D Kagome lattice, utilizing metallo-photonic waveguides \cite{Endo10},
photonic crystal structures \cite{Takida04}, or depositing a thin metal film 
\cite{Lai07,Kim11,Liew10,Kim08}. Flat bands can also be engineered in a frustrated lattice of micro-pillar optical cavities \cite{Baboux}. The lowest order Fourier decomposition of a Kagome
lattice potential $V_{ext}$ can be described using just three components,
written 
\begin{equation}
V_{ext}=V_{0}\left| f_{1}(\mathbf{x})e^{ik_{0}\mathbf{b}_{1}\cdot \mathbf{x}%
}+e^{ik_{0}\mathbf{b}_{2}\cdot \mathbf{x}}+e^{ik_{0}\mathbf{b}_{3}\cdot 
\mathbf{x}}\right| ^{2},  \label{eq18}
\end{equation}
where $f_{1}(\mathbf{x})=e^{ik_{0}px/(1+4p/3)}\cos (k_{0}px/(1+4p/3)),$ $%
\mathbf{b}_{1}=(1/(1+4p/3),0)$, $\mathbf{b}_{2}=(-1/(2+8p/3),-\sqrt{3}/2)$, $%
\mathbf{b}_{3}=(-1/(2+8p/3),\sqrt{3}/2)$, $V_{0}>0$ is the lattice
intensity, $k_{0}$ is a scalar quantity proportional to the lattice
constant, and $p\rightarrow 3/2$ \cite{Ablowitz09,Law09}. We implement the
above potential in the dGP equation Eq. (%
\ref{eq17}), and evolve forward in time starting from a resonantly induced vortex-antivortex
superposition state given by Eq. (\ref{eq21}).

\begin{figure}[tbp]
\includegraphics[width=\columnwidth]{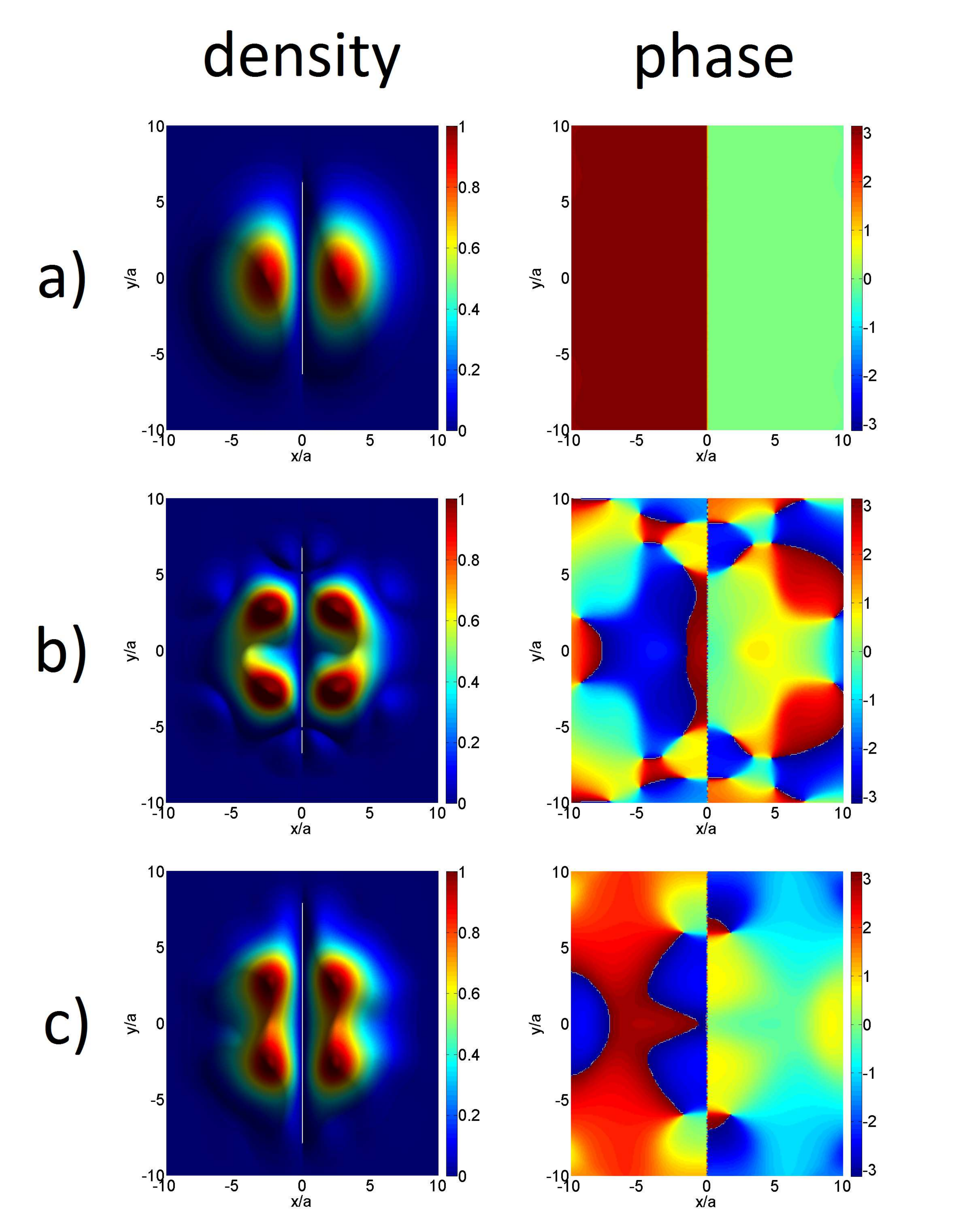}
\caption{Time evolution of a polariton BEC initially seeded with a vortex-antivortex superposition for a Kagome lattice potential given 
by Eq.  (\ref{eq18}). 
Timeframes for the  (a) initial state $t = 0$; (b) intermediate state $ t = 5 $; and (c) steady-state $ t = 10 $ are shown. Timescales are in units of $ \hbar/meV $. The following parameters were used in the simulations: $ \hbar g = 0.05$,  $ \hbar \gamma = 1.0$, and $ \hbar \eta = 0.1$, where energies are in $ meV $.}
\label{fig7}
\end{figure}

In Fig. \ref{fig7} we show the spatiotemporal evolution of the
vortex superposition state in a Kagome lattice potential.  Upon examination of Fig. \ref{fig7}, 
we can see that in the presence of a Kagome lattice, our vortex superposition state is coherent 
and stable as it is generated in the distributed Bragg reflector (DBR) microcavity outlined in Fig. \ref{fig1}.  
Fig. \ref{fig7} also illustrates the phase of the two-dimensional vortex superposition non-equilibrium 
condensate in a Kagome lattice potential, which shows the characteristic phase difference between the lobes of $\pi $.  It should be pointed out that
the flat band in the Kagome lattice corresponds to the third lowest energy band.  In our simulations the polaritons predominantly occupy the first band, hence the only effect of the bands is to periodically modulate the interference.  In order to target the flat band, techniques such as metastable condensation are required to create the condensate in these high momentum states.

\section{Sagnac Phase}

\label{sec:sagnac}

\subsection{Comparison to optical Sagnac interferometers}

Optical Mach-Zehnder interferometers \cite{Chow85} are typically capable of
achieving an excellent rotational measurement sensitivity, thanks to the
large area that the interferometric loop circumscribes \cite{Colella75,Riehle91}.
In this respect, atomic matter wave interferometers typically have
small loop areas and short-time rotational sensitivities \cite
{Gustavson97,McGuirk00}. However, matter wave interferometers have
the advantage that their de Broglie wavelength is much shorter than the
optical counterparts. For atomic BECs, this makes the rotational measurement sensitivity of
atomic matter wave interferometers per unit area exceed that of optical ones
by the ratio $mc^{2}/\hbar \omega \sim 10^{10}$ \cite{Dowling98}. Typically polariton masses
are extremely insignificant (i.e., $m_{\text{eff}} \sim 10^{-4} m_e $), where $m_e $ is the
bare electron mass. This almost entirely negates the advantage of polaritons
to $m_{\text{eff}} c^{2}/\hbar \omega \sim 1$, which is the same level as photons.
This is not surprising as polaritons themselves are part-matter, part-photon
quasiparticles, hence much of their characteristics are inherited from
light. In terms of sensitivity, polaritons would appear to be ill-suited for
such gyroscopic applications. We argue below that for counter-rotating vortex superposition based interferometers this argument does not directly apply, as the Sagnac phase is independent of the de Broglie wavelength. 

Let us first review the state-of-the-art optical Sagnac interferometers.  
Typically, in an optical interferometer, light of wavelength $
\lambda $ propagates in a closed loop. For each completed propagation in the
closed loop configuration, such as in a fiber optic gyroscope, the Sagnac phase is given by 
\begin{equation}
\phi_{\Omega }=\frac{8 \pi A_{\text{loop}}\Omega}{\lambda c},  \label{eq22}
\end{equation}
where the $\lambda $ is the wavelength of the light, $\Omega $ is the
angular (rotational) velocity, $c$ is the speed of light, and $A_{\text{loop} } $ is the area of the Mach-Zender interferometer. In a ring laser gyroscope, the quantity that is measured is the beat frequency of two counterpropagating lasers.  In terms of the phase difference, this is now a time-evolving quantity of the form
\begin{align}
\phi_{\Omega }=\frac{8 \pi A_{\text{loop}}\Omega t}{\lambda p},  \label{ringlaser}
\end{align}
where $ p $ is the perimeter of the interferometric loop. The fiber optic gyroscope and the ring laser gyroscope phases differ in operation in that the latter is time dependent, while the former is not. This means that as the system rotates, the fiber optic gyroscope  produces a shift in the fringes only when there is a rotation $ \Omega $, while the ring laser gyroscope detects the total time-varying phase of the rotation. Therefore the longer one waits between comparisons of the interference fringes, the larger the angle difference, and therefore the sensitivity.  

In both types of optical gyroscopes, the phase is proportional to the area of the interferometric loop.  
This is quite different to the situation in a BEC where the phase around a vortex is 
fixed purely by the topological winding number.  Thus independent of the path, a phase of $ 2 \pi l $ is 
picked up around a vortex.  For a fixed $ l $, the path can in principle enclose an arbitrary area and yet will 
not pick up any additional phase.  It is also independent of the de Broglie wavelength $ \lambda $ for the same reasons. 
Therefore, {\it for a counter-rotating vortex-antivortex superposition, the Sagnac phase is independent of the area of the BEC and the mass of the particles (i.e., polaritons) involved}.  As the Sagnac phase directly originates from the interference of two counter-rotating
angular momentum, it follows that the the Sagnac phase can then be written as 
\begin{equation}
\phi_{\Omega }(t)=2l\Omega t,  \label{eq24}
\end{equation}
and can be determined from the $\Omega $ of the vortex superposition state in a
polariton BEC.

Is it possible to regain the geometry and wavelength dependence of Eq. (\ref{eq22}) in a BEC to further improve the sensitivity? For the ring geometry considered in Sec. \ref{sec:meta} the area and wavelength dependence is reinstated, as we show below. For the BEC case the Sagnac
phase for polaritons can be written \cite{Thanvanthri12}
\begin{equation}
\phi _{\Omega }(t)=N(t)\frac{4m   A_{\text{loop} }    }{\hbar }\Omega .
\label{eq23}
\end{equation}
where $ N(t) $ is the number of times a polariton revolves around the loop in a time $ t $.  In a ring geometry of radius $ r $ as seen in \cite{Kalevich15,Dreismann}, we have for counter-propagating momenta $ \pm k_0 $
\begin{align}
N(t) = \frac{t\hbar k_0 }{2 \pi r m} .
\end{align}
This therefore gives
\begin{equation}
\phi_{\Omega }(t)=2 k_0 r \Omega  t    .
\label{ringphi}
\end{equation}
We note
here that each polariton in the BEC cloud has the same magnitude of the
angular momentum, which contributes to the signal-to-noise ratio as will be discussed below.  
The relation now has a boost in the sensitivity with the radius of the ring, which arises from the fact that the velocity of a wave with momentum $ k_0 $ has a value $ v = \hbar k_0 /m $.  This is in contrast to a vortex which decays with radius $ v = \hbar l /mr $.  

In the ring geometry we regain the enhancement due to the de Broglie wavelength in Eq. (\ref{ringphi}), as for heavy particles, the typical wavenumber $ k_0 = 2\pi/\lambda $ is much higher, which also results in enhanced sensitivities. In this sense, the use of periodic potentials as discussed in the previous section can aid the sensitivity, as this produces effectively larger effective masses, and hence larger typical $ k_0 $.  We note that
on setting $ k_0 = 2\pi/\lambda $, $ A_{\text{loop} } = \pi r^2 $, and $ p = 2 \pi r $, Eq. (\ref{ringphi}) reduces to Eq. (\ref{ringlaser}), so that BEC based gyroscopes can be considered to be equivalent to the ring laser gyroscopes up to geometrical factors and the wavelength of the particles being used.

\subsection{Detection of Sagnac phase}

When the laboratory frame rotates in time, the two counter-rotating matter
waves pick up different phases, and the transverse density profile at $z=0$
is given as 
\begin{equation}
\left( 1+\cos [2 l (\phi + \Omega t) ] \right) \left| \psi (%
\mathbf{x},t)\right| ^{2},  
\end{equation}
or for the case of the ring geometry,
\begin{equation}
\left( 1+\cos [ 2 k_0 r (\phi +\Omega t) ] \right) \left| \psi ( \mathbf{x},t)\right| ^{2}, 
\end{equation}
signifying a rotation of the interference pattern as the Sagnac phase accumulates over time.
Following the rotation of the lobes in the interference pattern, one may
determine the angular velocity $\Omega $ from the density profile of the
polariton BEC. The condensate density will show an
interference pattern determined by the phase difference between the
amplitudes and charges of the two vortex components, as seen in Figs. \ref{fig3old} - \ref{fig:ring}. This allows the polariton BEC to be used for the measurement of
changes in the Sagnac phase caused by the rotation of the laboratory frame
of reference (i.e., the interferometer itself).
This can be performed using existing experimental techniques where a CCD camera is used to image the photoluminescence from the polariton BEC. 

Determining the Sagnac phase using our proposed polariton
Sagnac interferometer will depend on the vortex charge $l$ of the polariton
BEC superposition state. In practice, spiral phase plates (SPPs) generally
work best for low integer values of $l$. However, in principle, $l$ can take on any integer value, although low integer values of $l$ generally give a more stable vortex superposition. The proposed
polariton Sagnac interferometer is based on imaging the standing wave of the
polariton BEC, which is comparably much smaller than the area occupied by comparable optical technologies (e.g., ring-laser gyroscope). Thus the polariton ring geometry has the potential for higher sensitivities.  

We are now in a position to compare the rotational phases per particle of the interference patterns $ \phi_{\Omega } $.  Due to the similar nature of the operation, it is simplest to compare the polariton gyroscopes to the ring laser gyroscope, which both have an increasing Sagnac phase in the time domain.  For the ring laser gyroscope, taking $A_{\text{loop}} \sim 1 \mbox{ m}^2 $ and $ p \sim 1 \mbox{ m}$, and typical optical wavelengths, one obtains the proportionality of $ \Omega t $ in Eq. (\ref{ringlaser}) to be $ 8 \pi A_{\text{loop}} 
/ \lambda p \sim 10^7 $.  For ultracold atom implementations, $A_{\text{loop}} \sim 1 \mbox{ mm}^2 $, $ p \sim 1 \mbox{ cm}$, and velocities $ v \sim 1 \mbox{m/s} $ we have $ 8 \pi A_{\text{loop}} / \lambda p \sim 10^6 $ \cite{Barrett14}.  For the polariton gyroscope with ring radii of $ r \sim 100 \mu $m and $ k_0 \sim 10 \mu \mbox{m}^{-1} $ \cite{Lai07}, this gives the coefficient of the rotational phase in Eq. (\ref{ringphi}) as $ k_0 r \sim 10^3 $.  Thus 
in terms of the best achievable sensitivities per particle, polariton gyroscopes appear to have a large handicap compared to other methods due to their long wavelengths and relatively small interferometric loop areas. However, we shall see in the next section that some of this deficit is made up by the superior signal-to-noise of polaritons.

\subsection{Signal-to-noise ratio and stability}

We now discuss the signal-to-noise ratio (SNR) and the sensitivity
achievable in such a polariton Sagnac interferometer. The SNR is given by
the ratio of the rotational phase shift to the shot-noise that depends on
the number of photons $N$ arriving at the CCD from the polariton condensate
per second \cite{Thanvanthri12,Zimmer} 
\begin{equation}
\text{SNR}=\phi _{\Omega }(t)/\sqrt{1/N}.  \label{eq28}
\end{equation}
The final SNR has a maximum at a certain rate of photon detection, and the
sensitivity $\Omega_{\min }$ is calculated by setting SNR$=1$. Thus, for the superposition in the geometries as seen in Figs. \ref{fig3old} - \ref{fig5}, 
\begin{align}
\Omega_{\min }=1/(2l t \sqrt{N}),
\label{regl} 
\end{align}
and for the vortex-antivortex superposition in the ring geometry of Fig. \ref{fig:ring} we have 
\begin{align}
\Omega_{\min }=1/(2k_0 r t \sqrt{N}). 
\end{align}
Typical polariton densities are in the region of 
$10^{9}$ cm$^{-2}$, and a polariton spot size of $100$ $\mu $m$^{2}$
gives a polariton number in the region of $10^{3}$. The lifetime of the polaritons is in the region of $ \sim 10 $ ps, hence we can estimate that the 
number of signal photons is $ N \sim 10^{14} \mbox{ s}^{-1} $ for polaritons.  This gives our estimate for the polariton vortex-antivortex superposition in the ring geometry
\begin{align}
\Omega_{\min } \sim  10^{-10} \mbox{ rad s}^{-1} \mbox{ Hz}^{-1/2} .
\end{align}
This is comparable with state-of-the-art atomic beam gyroscopes which are at the $ 10^{-10} \mbox{ rad s}^{-1} \mbox{ Hz}^{-1/2} $ level \cite{Gustavson97,Barrett14}.

With regard to the
stability of our proposed polariton Sagnac interferometer, it bypasses the
beam drift that is a common cause of instability in the metrology that
employs atomic or optical beams, and no complicated atomic beam
configurations are required.  The technical issues that optical
Sagnac interferometers face, such as the short-time rotational sensitivity
problem \cite{Wright08}, and the change in phase velocity of light \cite
{Stedman97}, should not be present for polaritons by construction. Moreover, thermal expansion over atomic beam
trajectories is not an issue for a polariton BEC, as atomic beams are not
needed. Interferometers which use atomic beams usually have beam drifts on
the order of $\mu \deg /h$ after hours of running time \cite{Durfee}. The simplicity of our scheme (i.e., no large optical loops in a solid state device), would suggest that our scheme should not suffer as seriously from stability issues.  

We note that for purposes of
seismometry, since the vector of angular velocity $\mathbf{\Omega }$ is
proportional to the vector of ground velocity $\mathbf{v}$ 
\begin{equation}
\mathbf{\Omega }=\frac{1}{2}\nabla \times \mathbf{v}.   \label{eq25}
\end{equation}
In this sense a quantum polariton seismometer can detect the ground velocity from the
Sagnac phase by detecting rotations that originate from unaccounted rotations of the ground.

\section{Summary and conclusion}

In this study, we have introduced two schemes for inducing counter-rotating vortex superposition
states in a polariton BEC for gyroscopy.  The first is via direct injection of orbital angular momentum
(OAM) of light through a distributed Bragg reflector (DBR) microcavity. Once
the vortex superposition state is induced, it follows a free evolution under
standard pump-loss dynamics of the polariton BEC. The second is via metastable condensation
in periodic lattices, where particular momentum modes are reinforced by using 
pumping profiles with the target periodicity. 
 By directly numerically
simulating the dynamics, we found that the vortex-antivortex superposition state is stable
under a variety of different conditions; in smooth, disordered, and periodic
potentials. In all cases we found that the vortex superposition state is stable and is long-lived.  

We investigated the possibility that the superposition of
counter-rotating currents in the polariton BEC can be used as a method for
determining the Sagnac phase -- and thus the angular velocity of the interferometer --
effectively constituting a ``quantum gyroscope.'' The vortex-antivortex superposition has the same Sagnac phase dependence as atomic BECs, as the phase relation around a vortex is topologically fixed, regardless of the particle mass or the area of the BEC. Optical Sagnac interferometers have large sensitivities
by increasing the area of the interference loop, and atomic Sagnac interferometers have an advantage due to their small de Broglie wavelength.  In this respect, polariton gyroscopes have relatively small loop areas and long wavelengths. This can be enhanced using ring geometries or periodic potentials.  In comparison with current technology it would appear that on a per polariton basis atomic and optical interferometers have a larger sensitivity. However, the polaritons do possess the great advantage that their signal-to-noise ratio should be considerably larger than that of atomic BECs.  In an atomic BEC the optical measurement has the limitation that it should not disturb the system \cite{Ilookeke14}, hence there are restrictions on how bright the probe light should be. In contrast, 
for polaritons they are themselves part light, hence illumination by laser light is intrinsic to the condensation process.  Thus while the phase dependence of the polariton Sagnac interferometer starts with several orders of magnitude deficit, the final estimated sensitivity is competitive with existing methods.  

Polariton BEC based gyroscopes have the additional practical advantage to atomic BECs that they should lead to compact and room temperature compatible devices, as it is based on semiconductor technology.  This could lead to a device with the desirable characteristics of compact operation with very high sensitivity and stability.  For the metastable condensation technique, the ability of electrical injection of the condensate would remove the need for optics entirely, simplifying the device further. Due to the limitations in the size of the polariton condensate, they are naturally suited towards small volume applications.  
The effective mass of the polaritons can be tuned by
engineering periodic structures, and conventional methods such as changing
the exciton fraction of the polaritons, which give additional boosts to sensitivity.

\begin{acknowledgments}
FIM would like to thank NASA Tech Briefs magazine and the Institution of Engineering and Technology (IET) for endorsing the Polariton Interferometer. JPD would like to acknowledge support from the AFOSR, the ARO, and the NSF. TB would like to acknowledge support from the Inamori foundation, NTT Basic Research Laboratories, and the Shanghai Research Challenge Fund.
We would like to gratefully acknowledge the interesting discussions with Bill Bischel at Gener8 Inc., Ram Yahalom at Infiber Technology Inc., and Ben Kora at Jacobs Engineering Group Inc. Finally, we thank Kathleen A. Bryan, Jidapa Chayakul, Bryan Gard, 
and Vincent Cellucci at the Louisiana State Univerisity College of Art \& Design, for their assistance with the artwork in Figure 1.
\end{acknowledgments}

\appendix

\section{Orbital Angular Momentum states of light}
\label{sec:app}

OAM superposition states of light can be generated by passing a TEM$_{00}$
beam mode through a spiral phase plate (SPP), followed by a Mach-Zehnder
interferometer with a Dove prism. The SPP converts the TEM$_{00}$ to a
Laguerre-Gaussian (i.e., LG$_{l,p}$) beam mode of quantized OAM with quantum
numbers $l$ and $p$, and the Dove prism changes the winding of the quantized
OAM from $l\hbar $ to $-l\hbar $. 
In addition to generation using spiral phase plates \cite{Turnbull96},
cylindrical lens mode converters \cite{Tamm90}, and computer generated
holograms \cite{Heckenberg92} are alternative methods to generate integer
quantized OAM. 
%
%
The LG$_{l,p}$ beam modes form a complete orthonormal basis set of solutions
for paraxial light beams commonly found in lasers. These modes can be
expressed as \cite{Allen} 
\begin{align}
u_{l,p}^{\text{LG}}& (r,\phi ,z)=\sqrt{\frac{2p!}{\pi w^{2}(z)(\left|
l\right| +p)!}}\left[ \frac{\sqrt{2}r}{w(z)}\right] ^{\left| l\right| }\exp
\left[ \frac{-r^{2}}{w^{2}(z)}\right]  \nonumber \\
& \times L_{p}^{\left| l\right| }\left( \frac{2r^{2}}{w^{2}(z)}\right) \exp
\left[ -i\left( \frac{kr^{2}}{2R(z)}+l\phi -\varphi (z)\right) \right] ,
\label{eq2}
\end{align}
where the radius of the beam squared is $w^{2}(z)=2(z^{2}+b^{2})/kb,$ the
radius of curvature is $R(z)=(z^{2}+b^{2})/z$, the Gouy phase $\varphi
(z)=(2p+\left| l\right| +1)\tan ^{-1}(z/b),$ $b$ is the Rayleigh range, 
$k$ is the wave number, and $%
L_{p}^{\left| l\right| }(r)$ describe the Laguerre polynomials, 
\begin{equation}
L_{p}^{\left| l\right| }(r)=\sum\limits_{m=0}^{p}(-1)^{m}\frac{(l+p)!}{%
(p-m)!(\left| l\right| +m)!m!}r^{m}.  \label{eq3}
\end{equation}
In Eqs. (\ref{eq2}) and (\ref{eq3}), $p$ is the number of nonaxial radial
nodes, and the index $l$ is known as the winding number. The winding number
describes the helical structure of the wavefront and the number of times the
phase jumps occur around the beam path in the azimuthal direction. When $%
l=p=0$, the LG$_{l,p}$ beam is identical to the fundamental Gaussian beam,
or TEM$_{00}.$

The higher order modes from the fundamental mode may be calculated using
the operator algebra formalism \cite{Stoler82}. Analogous to the two-dimensional
quantum harmonic oscillator \cite{Nienhuis,Messiah}, at $z=0$, 
\begin{equation}
\hat{A}_{x}(0)=\frac{1}{\sqrt{2bk}}(kr\cos (\phi )+b\cos (\phi )\partial
_{r}-br\sin (\phi )\partial _{\phi }),  \label{eq4}
\end{equation}
and 
\begin{equation}
\hat{A}_{y}(0)=\frac{1}{\sqrt{2bk}}(kr\sin (\phi )+b\sin (\phi )\partial
_{r}+br\cos (\phi )\partial _{\phi }).  \label{eq5}
\end{equation}
These operators can be evolved in the $z$-direction by a propagator $\hat{U}%
(z)$ for $\partial _{\phi }\partial _{r}=\partial _{r}\partial _{\phi }$
according to 
\begin{eqnarray}
\hat{U}(z) &=&\exp \left[ -\frac{i}{2k}\hat{P^{2}}z\right]  \nonumber \\
&=&\exp \left[ -\frac{i}{2k}(\partial _{r}^{2}+r^{2}\partial _{\phi
}^{2})z\right] .  \label{eq6}
\end{eqnarray}
Next, we can rewrite Eqs. (\ref{eq4}) and (\ref{eq5}) as 
\begin{equation}
\hat{A}_{x}(z)=\hat{U}(z)\hat{A}_{x}(0)\hat{U}^{\dagger }(z),  \label{eq7}
\end{equation}
\begin{equation}
\hat{A}_{y}(z)=\hat{U}(z)\hat{A}_{y}(0)\hat{U}^{\dagger }(z),  \label{eq8}
\end{equation}
where 
\begin{eqnarray}
\hat{A}_{\pm }(z) &=&\frac{1}{\sqrt{2}}\left[ \hat{A}_{x}(z)\mp i\hat{A}%
_{y}(z)\right]  \nonumber  \label{eq9} \\
&=&\frac{1}{2\sqrt{bk}}\left[ kre^{\mp i\phi }+(be^{\mp i\phi }+2ze^{\pm
i\phi })(\partial _{r}+r\partial _{\phi })\right] .  \nonumber \\
&&
\end{eqnarray}
The operators in Eq. (\ref{eq9}) obey the commutation rules 
\begin{equation}
\left[ \hat{A}_{\pm }(z),\hat{A}_{\pm }^{\dagger }(z)\right] =1,\left[ \hat{A%
}_{\pm }(z),\hat{A}_{\mp }^{\dagger }(z)\right] =0.  \label{eq10}
\end{equation}
Higher-order LG$_{l,p}$ beam modes can be obtained from operating on the TEM$%
_{00}$ according to \cite{Nienhuis} 
\begin{equation}
u_{l,p}^{\text{LG}}(z)=\sqrt{\frac{1}{m!n!}}\left[ \hat{A}_{-}^{\dagger
}(z)\right] ^{n}\left[ \hat{A}_{+}^{\dagger }(z)\right] ^{m}\text{TEM}%
_{00}(z),  \label{eq11}
\end{equation}
where $m=(l+p)/2$ and $n=(l-p)/2.$ We note that OAM
states $u_{l,p}^{\text{LG}}(z)$ and $u_{-l,p}^{\text{LG}}(z)$ differ only in
the direction of the phase winding, that is, clockwise or counterclockwise,
respectively. 


\begin{thebibliography}{99}
\bibitem{Ketterle}  Ketterle W, Andrews M R, Davis K B, Durfee D S, Kurn D
M, Mewes M O and Van Druten N J 1996 \textit{Phys. Scr.} \textbf{T66} 31

\bibitem{Anderson}  Anderson M H, Ensher J R, Matthews M R, Wieman C E and
Cornell E A 1995 \textit{Science} 269 \textbf{5221} 198

\bibitem{Bloch}  Bloch I, Dalibard J and Nascimb\`{e}ne S 2012\ \textit{%
Nature Phys.} \textbf{8} 267

\bibitem{Byrnes12}  Byrnes T, Wen K and Yamamoto Y 2012 \textit{Phys. Rev. A}
\textbf{85} 040306

\bibitem{Byrnes13}  Byrnes T 2013 \textit{Phys. Rev. A} \textbf{88} 023609

\bibitem{Schumm}  Schumm T, Hofferberth S, Andersson L M, Wildermuth S,
Groth S, Bar-Joseph I, Schmiedmayer J and Kr\"{u}ger P 2005 \textit{Nature
Phys.} \textbf{1} 57

\bibitem{Deng}  Deng H, Haug H and Yamamoto Y\ 2010 \textit{Rev. Mod. Phys.} 
\textbf{82} 1489

\bibitem{Byrnes14}  Byrnes T, Kim N Y and Y. Yamamoto 2014 \textit{Nature
Phys.} \textbf{10} 803

\bibitem{Byrnes132}  Byrnes T, Yamamoto Y and Loock P V 2013\ \textit{Phys.
Rev. B} \textbf{87} 201301

\bibitem{Plumhof}  Plumhof J D, St\"{o}ferle T, Mai L, Scherf U and Mahrt R
F 2014 \textit{Nature Mater.} \textbf{13} 247

\bibitem{Masumoto}  Masumoto N, Kim N Y, Byrnes T, Kusudo K, L\"{o}ffler A, H%
\"{o}fling S, Forchel A and Yamamoto Y 2012 \textit{New J. Phys.} \textbf{14}
065002

\bibitem{Jacqmin}  Jacqmin T, Carusotto I, Sagnes I, Abbarchi M, Solnyshkov
D D, Malpuech G, Galopin E, Lema\^{i}tre A, Bloch J and Amo A 2014\ \textit{%
Phys. Rev. Lett.} \textbf{112} 116402

\bibitem{Houdre}  Houdr\'{e} R., Stanley R P, Oesterle U, Ilegems M and
Weisbuch C 1994\ \textit{Phys. Rev. B} \textbf{49} 16761

\bibitem{Ge13}  Ge L, Nersisyan A, Oztop B and T\"{u}reci H E  2013\ \textit{arXiv preprint} arXiv:1311.4847

\bibitem{Keeling}  Keeling J and Berloff N G 2011\ \textit{Contemp. Phys.} 
\textbf{52} 131

\bibitem{Lesne}  Lesne A and Lagu\"{e}s M 2011 \textit{SSBM }

\bibitem{Medard}  M\'{e}dard F, Trichet A, Chen Z, Dang L S and Richard M
2013\ \textit{Phys. Quantum Fluids} \textbf{177} 231

\bibitem{Franchetti12}  Franchetti G, Berloff N G and Baumberg J J 2012\ 
\textit{arXiv preprint} arXiv:1210.1187 

\bibitem{Clarke}  Clarke J 1994 \textit{Sci. Am.} \textbf{271} 46

\bibitem{Riedel}  Max R F, B\"{o}hi P, Li Y, H\"{a}nsch T W, Sinatra A and
Treutlein P 2010 \textit{Nature} \textbf{464} 1170

\bibitem{Okulov} Okulov A 2013 \textit{J. Low Temp. Phys.} \textbf{171} 397

\bibitem{Maucher}  Maucher F, Henkel N, Saffman M, Kr\'{o}likowski W, Skupin
S and Pohl T 2011\ \textit{Phys. Rev. Lett.} \textbf{106} 170401

\bibitem{Yin}  Yin C, Berloff N G, P\'{e}rez-Garc\'{i}a V M, Novoa D,
Carpentier A V and Michinel H 2011\ \textit{Phys. Rev. A} \textbf{83 }051605

\bibitem{Gu} Gu L, Huang H and Gan Z 2011 \textit{Phys. Rev. B} \textbf{84} 075402

\bibitem{Sturm} Sturm C, Tanese D, Nguyen H S, Flayac H, Galopin E, Lemaître A, Sagnes I et al. 2014 
\textit{Nature Commun.} \textbf{5} 3278

\bibitem{Lerario}  Lerario G, Cannavale A, Ballarini D, Dominici L, Giorgi M
D, Liscidini M, Gerace D, Sanvitto D and Gigli G 2014\ \textit{Opt. Lett.} 
\textbf{39} 2068

\bibitem{Dall}  Dall R, Fraser M D, Desyatnikov A S, Li G, Brodbeck S, Kamp
M, Schneider C, H\"{o}fling S and Ostrovskaya E A 2014\ \textit{Phys. Rev.
Lett.} \textbf{113} 200404

\bibitem{Gross}  Gross C 2012\ \textit{J. Phys. B At. Mol. Opt. Phys.} 
\textbf{45} 103001

\bibitem{Borde}  Bord\'{e} Ch J 2002\ \textit{Metrologia} \textbf{39} 435

\bibitem{Michel}  Michel R, Ampuero J, Avouac J, Lapusta N, Leprince S,
Redding D C and Somala S N 2013 \textit{IEEE TGRS} \textbf{51} 695

\bibitem{Dickerson}  Dickerson S M, Hogan J M, Sugarbaker A, Johnson D M S
and Kasevich M A 2013\ \textit{Phys. Rev. Lett.} \textbf{111} 083001

\bibitem{Sun} Sun K, Fejer M M, Gustafson E and Byer R L 1996 \textit{Phys. Rev. Lett.} 
\textbf{76} 3053

\bibitem{Gasparini}  Gasparini M A 2005\ \textit{Phys. Rev. D} \textbf{72}
104012

\bibitem{Edwards}  Edwards M 2013\ \textit{Nature Phys.} \textbf{9} 68

\bibitem{Delgado} Delgado A, Schleich W P and Süssmann G 2002 \textit{New J. Phys.} 
\textbf{4} 37

\bibitem{Dowling98}  Dowling J P 1998\ \textit{Phys. Rev. A} \textbf{57} 4736

\bibitem{Nandi}  Nandi G, Walser R and Schleich W P 2004\ \textit{Phys. Rev.
A} \textbf{69} 063606

\bibitem{Roumpos}  Roumpos G, Fraser M D, L\"{o}ffler A, H\"{o}fling S,
Forchel A and Yamamoto Y 2011 \textit{Nature Phys. }\textbf{7} 129

\bibitem{Grosso}  Grosso G, Nardin G, Morier-Genoud F, L\'{e}ger Y and
Deveaud-Pl\'{e}dran B 2011 \textit{Phys. Rev. Lett.} \textbf{107} 245301

\bibitem{Molina-Terriza}  Molina-Terriza G, Torres J P and Torner L 2007\ 
\textit{Nature Phys.} \textbf{3} 305

\bibitem{Yakimenko}  Yakimenko A I, Bidasyuk Y M, Prikhodko O O, Vilchinskii
S I, Ostrovskaya E A and Kivshar Y S 2013 \textit{Phys. Rev. A} \textbf{88}
043637

\bibitem{Krizhanovskii}  Krizhanovskii D N, Whittaker D M, Bradley R A, Guda
K, Sarkar D, Sanvitto D, Vina L et al. 2010 \textit{Phys. Rev. Lett.} 
\textbf{104} 126402

\bibitem{Sanvitto10}  Sanvitto D, Marchetti F M, Szyma\'{n}ska M H, Tosi G, Baudisch M, Laussy F P, Krizhanovskii D N et al. 2010 \textit{Nat. Phys.} \textbf{6} 527

\bibitem{Marzlin}  Marzlin K P, Zhang W and Wright E M 1997\ \textit{Phys.
Rev. Lett.} \textbf{79} 4728

\bibitem{Wright09}  Wright K C, Leslie L S, Hansen A and Bigelow N P 2009\ 
\textit{Phys. Rev. Lett.} \textbf{102} 030405

\bibitem{Dreismann}  Dreismann A, Cristofolini P, Balili R, Christmann G,
Pinsker F, Berloff N G, Hatzopoulos Z, Savvidis P G and Baumberg JJ 2014 
\textit{Proc. Natl. Acad. Sci. U.S.A.} \textbf{111} 8770

\bibitem{Tosi}  Tosi G, Marchetti F M, Sanvitto D, Ant\'{o}n C, Szymanska M
H, Berceanu A, Tejedor C et al. 2011 \textit{Phys. Rev. Lett.} \textbf{107}
036401

\bibitem{Lagoudakis}  Lagoudakis K G, Wouters M, Richard M, Baas A,
Carusotto I, Andr\'{e} R, Dang L S and Deveaud-Pl\'{e}dran B 2008 \textit{%
Nature Phys.} \textbf{4} 706

\bibitem{Wouters07}  Wouters M and Carusotto I 2007\ \textit{Phys. Rev. Lett.%
} \textbf{99} 140402

\bibitem{Wouters10}  Wouters M and Carusotto I 2010\ \textit{Phys. Rev. Lett.%
} \textbf{105} 020602

\bibitem{Moxley15}  Moxley F I, Byrnes T, Ma B, Yan Y and Dai W 2015\ 
\textit{J. Comput. Phys.} \textbf{282} 303

\bibitem{Moxley13}  Moxley F I, Chuss D T and Dai W 2013\ \textit{Comput.
Phys. Commun.} \textbf{184} 1834

\bibitem{Pitaevskii}  Pitaevskii L P, Stringari S (Clarendon Press, 2003) 
\textit{Bose-Einstein Condensation} 

\bibitem{Manni11}  Manni F, Lagoudakis K G, Liew T C H, Andr\'{e} R and Deveaud-Pl\'{e}dran B  2011\ 
\textit{Phys. Rev. Lett.} \textbf{107} 106401

\bibitem{Wouters102}  Wouters M, Liew T and Savona V 2010\ 
\textit{Phys. Rev. B} \textbf{82} 245315

\bibitem{Fickler12}  Fickler R, Lapkiewicz R, Plick W N, Krenn M, Schaeff C, Ramelow S and Zeilinger A 2012\ 
\textit{Science} \textbf{338} 640

\bibitem{Lai07}  Lai C W, Kim N Y, Utsunomiya S, Roumpos G, Deng H, Fraser M
D, Byrnes T et al. 2007\ \textit{Nature} \textbf{450} 529

\bibitem{Kim08}  Kim N Y, Lai C W, Utsunomiya S, Roumpos G, Fraser M, Deng
H, Byrnes T et al. 2008\ \textit{Phys. Status Solidi B} 245 1076

\bibitem{Mekata03}  Mekata M 2003\ \textit{Phys. Today} \textbf{56} 12

\bibitem{Levi07}  Levi B G 2007 \textit{Phys. Today} \textbf{60} 16

\bibitem{Bergman08}  Bergmann K, Theuer H and Shore B W 1998 \textit{Rev.
Mod. Phys.} \textbf{70} 1003

\bibitem{Endo10}  Endo S, Oka T and Aoki H 2010 \textit{Phys. Rev. B} 
\textbf{81} 113104

\bibitem{Takida04}  Takeda H, Takashima T and Yoshino K 2004\ \textit{J.
Phys. Condens. Matter.} \textbf{16} 6317

\bibitem{Kim11}  Kim N Y, Kusudo K, Wu C, Masumoto N, L\"{o}ffler A, H\"{o}%
fling S, Kumada N, Worschech L, Forchel A and Yamamoto Y 2011\ \textit{%
Nature Phys.} \textbf{7} 681

\bibitem{Liew10}  Liew T C H, Kavokin A V, Ostatnick\'{y} T, Kaliteevski M,
Shelykh I A and Abram R A 2010\ \textit{Phys. Rev. B} \textbf{82} 033302

\bibitem{Baboux}  Baboux F, Ge L, Jacqmin T, Biondi M, Lema\^{i}tre A, Le
Gratiet L, Sagnes I et al. 2015 \textit{arXiv preprint} arXiv:1505.05652

\bibitem{Ablowitz09}  Ablowitz M J, Nixon S D and Zhu Y 2009\ \textit{Phys.
Rev. A} \textbf{79} 053830

\bibitem{Law09}  Law K J H, Saxena A, Kevrekidis P G and Bishop A R 2009\ 
\textit{Phys. Rev. A} \textbf{79} 053818

\bibitem{Chow85}  Chow W W, Gea-Banacloche J, Pedrotti L M, Sanders V E,
Schleich W and Scully M O 1985\ \textit{Rev. Mod. Phys.} \textbf{57} 61

\bibitem{Colella75}  Colella R, Overhauser A W and Werner S A 1975\ \textit{%
Phys. Rev. Lett.} \textbf{34} 1472

\bibitem{Riehle91}  Riehle F, Kisters Th, Witte A, Helmcke J and Bord\'{e}
Ch J 1991\ \textit{Phys. Rev. Lett.} \textbf{67} 177

\bibitem{Gustavson97}  Gustavson T L, Bouyer P and Kasevich M A 1997\ 
\textit{Phys. Rev. Lett.} \textbf{78} 2046

\bibitem{McGuirk00}  McGuirk J M, Snadden M J and Kasevich M A 2000\ \textit{%
Phys. Rev. Lett.} \textbf{85} 4498

\bibitem{Thanvanthri12}  Thanvanthri S, Kapale K T and Dowling J P 2012\ 
\textit{J. Mod. Opt.} \textbf{59} 1180

\bibitem{Kalevich15}  Kalevich V K, Afanasiev M M, Lukoshkin V A, Solnyshkov D D, Malpuech G, Kavokin K V, Tsintzos S I, Hatzopoulos Z, Savvidis P G and Kavokin A V 2015\ \textit{Phys. Rev.
B} \textbf{91} 045305

\bibitem{Barrett14}
Barrett B et al. 2014 \textit{C. R. Physique} \textbf{15} 875 

\bibitem{Zimmer}  Zimmer F E and Fleischhauer M 2006\ \textit{Phys. Rev. A} 
\textbf{74} 063609

\bibitem{Wright08}  Wright K C, Leslie L S and Bigelow N P 2008\ \textit{%
Phys. Rev. A} \textbf{77} 041601

\bibitem{Stedman97}  Stedman G E 1997\ \textit{Rep. Prog. Phys.} \textbf{60}
615

\bibitem{Durfee}  Durfee D S, Shaham Y K and Kasevich M A 2006\ \textit{%
Phys. Rev. Lett.} \textbf{97} 240801

\bibitem{Ilookeke14}
Ilo-Okeke E and Byrnes T 2014 \textit{Phys. Rev. Lett.}  \textbf{112} 233602

\bibitem{Turnbull96}  Turnbull G A, Robertson D A, Smith G M, Allen L and
Padgett M J 1996\ \textit{Opt. Commun.} \textbf{127} 183

\bibitem{Tamm90}  Tamm Chr and Weiss C O 1990\ \textit{JOSA B} \textbf{7}
1034

\bibitem{Heckenberg92}  Heckenberg N R, McDuff R, Smith C P and White A G
1992\ \textit{Opt. Lett.} \textbf{17} 221

\bibitem{Allen}  Allen L, Beijersbergen M W, Spreeuw R J C and Woerdman J P
1992\ \textit{Phys. Rev. A} \textbf{45} 8185

\bibitem{Stoler82}  Stoler D 1981 \textit{JOSA} \textbf{71} 334

\bibitem{Nienhuis}  Nienhuis G and Allen L 1993\ \textit{Phys. Rev. A} 
\textbf{48} 656

\bibitem{Messiah}  Messiah A 1995 M\'{e}canique Quantique \textbf{2} \textit{%
Dunod}

\end{thebibliography}

\end{document}